\documentclass[preprint,12pt]{elsarticle}

\usepackage{graphicx}
\usepackage{subcaption}
\usepackage{amssymb}
\usepackage{bm}
\usepackage{amsmath}
\usepackage[hidelinks]{hyperref}
\hypersetup{
  colorlinks   = true, %Colours links instead of ugly boxes
  urlcolor     = blue, %Colour for external hyperlinks
  linkcolor    = blue, %Colour of internal links
  citecolor   = red %Colour of citations
}

\newcounter{bla}

\journal{Computer Physics Communications}

\begin{document}

\begin{frontmatter}



\title{OpenDust: A fast GPU-accelerated code for calculation forces, acting on microparticles in a plasma flow }


\author[a,b]{D. Kolotinskii\corref{author}}
\author[a,b,c]{A.Timofeev}

\cortext[author] {Corresponding author.\\\textit{E-mail address:} kolotinskiy.da@phystech.edu}
\address[a]{Moscow Institute of Physics and Technology, Dolgoprudnyi, Moscow region, 141701, Russia}
\address[b]{Joint Institute for High Temperatures, Russian Academy of Sciences, Moscow, 125412, Russia}
\address[c]{HSE University, Moscow, 101000, Russia}

\begin{abstract}
We present the first open-source, GPU-based code for complex plasmas. The code, OpenDust, aims to provide researchers both experimenters and theorists user-friendly and high-performance tool for self-consistent calculation forces, acting on microparticles, and microparticles' charges in a plasma flow. OpenDust performance originates from highly-optimized Cuda back-end and allows to perform self-consistent calculation of plasma flow around microparticles in seconds. This code outperforms all available codes for self-consistent complex plasma simulation. Moreover, OpenDust can also be used for simulation of larger systems of dust microparticles, that was unavailable before. OpenDust interface is written in Python, which  provides ease-of-use and simple installation from Conda repository. 

\end{abstract}

\begin{keyword}
Python; GPU; Dusty plasma; Complex plasma; Ion wake; OpenMM.

\end{keyword}

\end{frontmatter}

% \linenumbers

% All CPiP articles must contain the following
% PROGRAM SUMMARY.

{\bf PROGRAM SUMMARY}
  %Delete as appropriate.

\begin{small}
\noindent
{\em Program Title:} OpenDust                                        \\
{\em CPC Library link to program files:} (to be added by Technical Editor) \\
{\em Developer's repository link:} \href{https://github.com/kolotinsky1998/opendust}{https://github.com/kolotinsky1998/opendust} \\
{\em Code Ocean capsule:} (to be added by Technical Editor)\\
{\em Licensing provisions:} MIT  \\
{\em Programming language:} Python                                  \\
{\em Nature of problem:} GPU cards can significantly speedup self-consistent calculations of forces, acting on microparticles in a plasma flow. The available codes use CPUs or not compiled into libraries or packages which can be used. Therefore, researchers need to spend much time writing their own codes or use less effective ones. \\

{\em Solution method:}
Development of a highly-optimized GPU-accelerated library for self-consistent simulations of streaming plasma around microparticles. The library's interface is entirely written in Python for enhanced user- friendliness.

\end{small}

\section{Introduction}

Problem of calculation forces, acting on microparticles of condensed matter, so-called dust particles, in a plasma flowing environment, arises in a variety of industry applications \cite{krainov2020dielectric,nguyen2020formation, shoyama2021charging}  and fundamental issues \cite{ivlev2015statistical, nikolaev2021nonhomogeneity, lisin2020experimental, kong2011one}, related to the complex plasma physics \cite{fortov2005complex}. For example, flowing plasma can lead to the release of pollutant particles from a processed sample during extreme ultraviolet lithography important for microelectronics \cite{krainov2020dielectric, shoyama2021charging, nguyen2020formation}, which can lower the quality and productivity of the manufacturing processes. Controlling and minimizing contamination of such particles requires detailed study of the forces, acting on them. Besides, challenge of dust particles managing arises in the fields of controlled nuclear fusion and thin film deposition, for which calculation forces, acting on dust particles, is necessary \cite{fortov2005complex}. Understanding of mechanisms behind plasma-particle interactions is necessary to explain various experimentally observed extraordinary phenomenon: nonreciprocal effective interaction between microparticles suspended in a radio‐frequency produced plasma sheath \cite{ivlev2015statistical,lisin2020experimental,ignatov2019collective}, formation of chain-like structures of microparticles in a plasma flow \cite{kong2011one,polyakov2011structural}, non-homogeneity of phase state in a complex plasma mono-layer \cite{nikolaev2021nonhomogeneity,klumov2019effect}.

\begin{table}[ht!]
  \begin{center}
    \caption{The available capabilities of OpenDust version 1.0.0.}
    \label{tab:table1}
    \begin{tabular}{p{0.4\textwidth}|p{0.5\textwidth}}
      \hline
      \raggedright Options & \raggedright Features \tabularnewline 
      \hline
      \raggedright electron treatment & Boltzmann fluid \tabularnewline 
      \hline
      \raggedright ion-neutral collisions models & \raggedright collisionless, resonant charge exchange collisions with constant frequency \tabularnewline 
      \hline
      \raggedright boundary conditions & \raggedright open boundary \tabularnewline
      \hline
      \raggedright dust particle charging models & \raggedright constant charge, orbital motion limited charge calculation, orbital motion limited electron flux and self-consistent ion flux calculation \tabularnewline
      \hline
      \raggedright simulation domain geometries & \raggedright cylinder \tabularnewline
      \hline
      \raggedright observables &  \raggedright time-dependent dust particle charges, time-dependent forces, acting on dust particles, trajectory of ions \tabularnewline
      \hline
      
    \end{tabular}
  \end{center}
 \end{table}

 Immersed to a plasma flow, dust particles typically gain large negative charge $\sim 10^4e$ \cite{fortov2005complex} and interact with each other and charged plasma species. Therefore, calculation of forces, acting on dust particles in a plasma environment, requires detailed description of the neighboring distribution of plasma species \cite{hutchinson2012intergrain}, which is inaccessible for the current experimental measurement techniques. Analytical approximations of forces, acting on dust particles, are restricted and can be used only in simple cases \cite{khrapak2002ion}. That leads to necessity of numerical simulations of plasma dynamics in the presence of dust particles. Such simulations are resource intensive and requires high optimized algorithms to compute forces in reasonable time \cite{hutchinson2007computation}.

 Historically, complex plasma physicists have developed their own codes, because of the specificity of the issues arising in that field. First numerical models were based on the Monte-Carlo simulation of ion distribution around dust particles \cite{schweigert1996alignment}. Hereafter, the Particle-In-Cell approach for simulation of complex plasma systems were actively developed \cite{hutchinson2002ion, hutchinson2011nonlinear, miloch2010wake, lampe2015grain, sukhinin2017plasma}. Recently, a new method of calculation force on dust particles in plasma environment were proposed by Alexander Piel \cite{piel2017molecular}. This method is based on GPU-accelerated molecular dynamics and allowed for the first time to simulate motion of two dust particle consistently with calculation of plasma dynamics \cite{matthews2020dust}. In spite of contemporary development in complex plasma computational methods, there is still no computationally effective open-source tool, which allows researchers to calculate force, acting on dust particles in a plasma flowing environment. 
 
Here, we present OpenDust, a fast code for self-consistent calculation forces, acting on dust particles, immersed in a flow of weakly-coupled classic plasma. OpenDust is based on molecular dynamics approach for plasma simulation \cite{piel2017molecular, vladimirov2003molecular, silvestri2022sarkas} and written in Python programming languages. Its high-performance originates from using GPU-accelerated library for molecular simulation Open\-MM \cite{eastman2017openmm}. OpenMM is used for acceleration the most resource intensive part of simulation, dynamics of plasma species.  OpenDust outperforms all available codes for self-consistent complex plasma simulation. Moreover, OpenDust can also be used for simulation of larger systems of dust microparticles, that was unavailable before. OpenDust aims at lowering the entry barrier for complex plasma simulations and offers user friendly Python-interface. In the Section 2, methods used in OpenDust for a complex plasma system simulation are presented. Code capabilities and structure are described in the Section 3. In the Section 4, test cases simulated with OpenDust are shown. Performance benchmarks of the code are presented in the Section 5.

\section{Methods}
\begin{figure}[hbt]
    \centering
    \includegraphics[width=0.8\textwidth, clip]{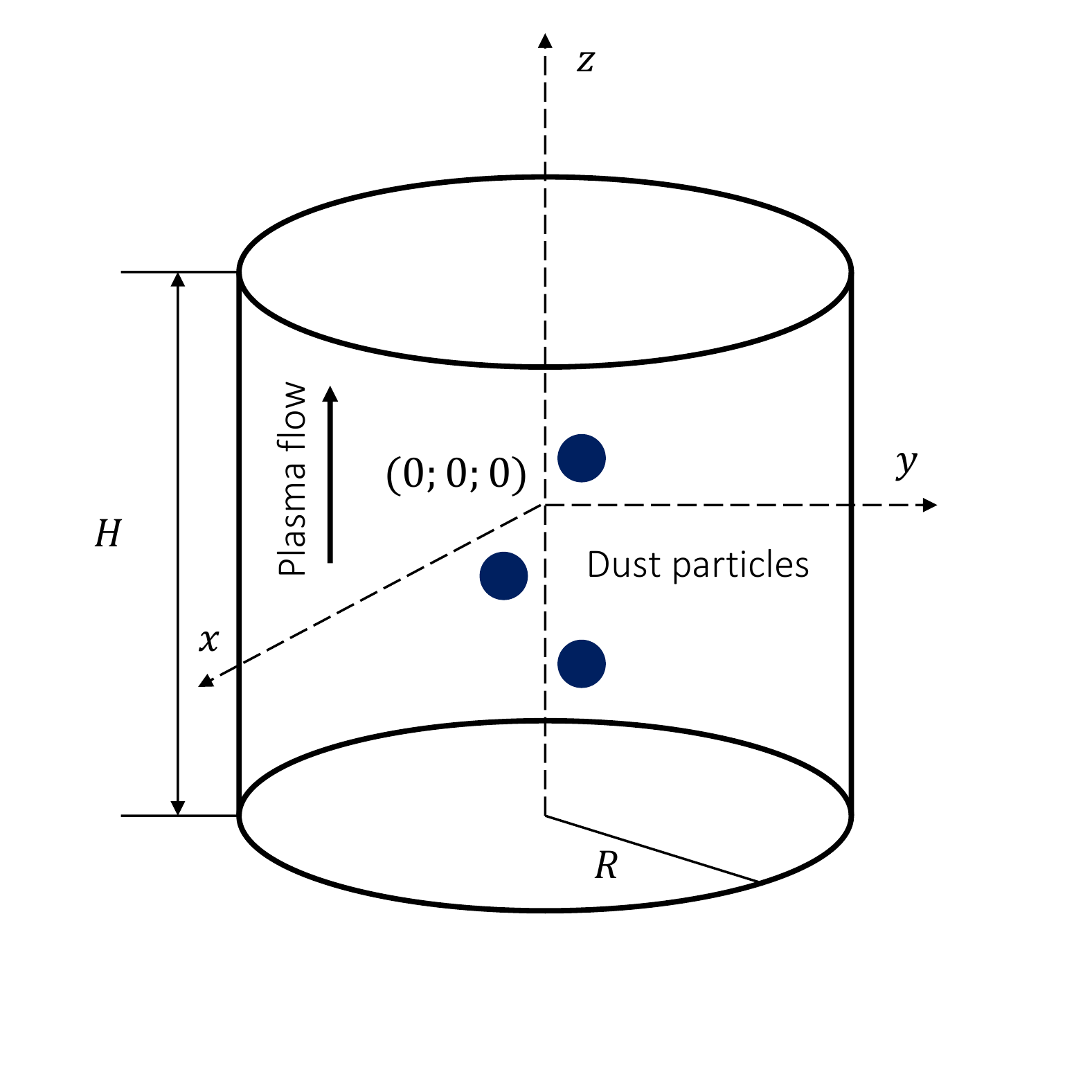}
    \caption{\label{fig:domain} Illustration of the simulation domain geometry used in OpenDust.}
\end{figure}

OpenDust operates in 3D Cartesian coordinates with a cylindrical computational domain filled with plasma, in which a system of dust particles can be simulated. The illustration of the simulation domain geometry is shown in the Figure \ref{fig:domain}. $H, R$ denote height and cylinder radius correspondingly. The applicate axis is parallel to the cylinder axis and is directed to the top side of the cylinder. The origin is located on the cylinder axis at the distance $\frac{H}{2}$ from the cylinder's bottom. Plasma flow if given is coaxial to the applicate axis.   

Streaming ions population is described as discrete particles with the same charge-to-mass ratio as ions. Charge of such discrete particles may be up to several hundred ion charges. The trajectory of motion of these discrete particles in a given electric field is the same as for real ions. Such discrete particles approach is used to reduce computational cost of calculations and is widely spread in plasma simulations \cite{hutchinson2012intergrain, piel2017molecular}. Further in this section, we use the term ion considering the described discrete particle. Electrons are treated as fluid governed by the Boltzmann factor $\exp{\left(e \Phi /k_B T_e\right)}$, where $e$ is the elementary electric charge, $k_B$ is the Boltzmann constant, $T_e$ is the electron temperature, $\Phi$ is the electric potential. Two limiting cases can be identified. Far from the charged dust particles, $|e\Phi| \ll k_B T_e$, and a linear approximation of the Poisson–Boltzmann equation lead to shielded Coulomb potentials for the ions. Very close to a dust particle, $|e\Phi| \gg k_B T_e$, and the small electron density allows them to be neglected, resulting in bare Coulomb potentials for the dust and ions. Following the Piel's approach \cite{piel2017molecular}, we use asymmetric model for dust-ion interactions to address the continuum transition between these two limiting cases. The ion-ion interactions are treated as shielded Coulomb interactions, while the force on the ions from the dust arises from the dust Coulomb potential. Ions and dust particles in the computational domain are subject to a confinement force from the assumed infinite homogeneous distribution of ions outside the simulation region. The approach for considering this force is described in details in Appendix A.

Two types of the ion flow are available at the moment in OpenDust. The first one is a collisionless flow, in which ions have the shifted Maxwellian velocity distribution \cite{hutchinson2011nonlinear}. The ion trajectories are calculated with the velocity Verlet algorithm \cite{scott1991computer}. The second one is a field driven flow of ions, colliding with neutral molecules \cite{kompaneets2016wakes, sundar2017impact}. The driving electric field is homogeneous and coaxial to the applicate axis. Collisions are characterized by constant frequency and are incorporated to simulate resonant charge exchange collisions with neutral molecules. The ion trajectories are calculated with the velocity Verlet algorithm \cite{scott1991computer}. Collisions are modeled using the Anderson thermostat \cite{andersen1980molecular}.

Open boundary conditions are used in OpenDust to simulate the ion flow through the computational domain. Any ion, that leaves the computational domain or is absorbed by dust particles, is replaced by a newly injected ion at a random position on the cylindrical domain boundary using a modified flux-conserving algorithm \cite{hutchinson2002ion}. These boundary conditions assume that the ions, entering the computational domain, are distributed to mimic ion distribution function in a region of homogeneous stationary plasma. Because of this assumption simulated dust particles should be located far enough from the domain boundary to their influence on the boundary ions can be neglected. The used boundary conditions are in details described in Appendix B.

In OpenDust, forces, acting on dust particles, consist of four parts.
\begin{enumerate}

\item The electric force between dusts and shielded ions, including in-domain and out-domain terms
\begin{equation}
\begin{aligned}
\boldsymbol{F_{id}^o} = \frac{q_i Q_d}{4\pi\varepsilon_0}\sum_{k}\frac{1+|\boldsymbol{r_i^k}-\boldsymbol{r_d}|/r_{D_e}}{|\boldsymbol{r_i^k}-\boldsymbol{r_d}|^3}\exp{\left(-\frac{|\boldsymbol{r_i^k}-\boldsymbol{r_d}|}{r_{D_e}}\right)}(  \boldsymbol{r_d}-\boldsymbol{r_i^k})\\ + \boldsymbol{E_{out}} Q_d,
\label{eq:Fido}
\end{aligned}
\end{equation}
where $\varepsilon_0$ is the vacuum permittivity, $\boldsymbol{r_i^k}$ is the ion radius vector, $\boldsymbol{r_d}$ is the dust radius vector, $r_{D_e}$ is the electron Debye radius, $Q_d$ is the dust particle charge, $q_i$ is the ion charge, $\boldsymbol{E_{out}}$ is the electric field from infinite homogeneously distributed ions outside the simulation region. Summation is carried out over all ions in the computational domain. 
\item The Coulomb force between dust particles
\begin{equation}
\boldsymbol{F_{dd^k}} = \frac{1}{4\pi\varepsilon_0}\sum_{p \neq k } Q_d ^k Q_d^p \frac{(  \boldsymbol{r_d^k}-\boldsymbol{r_d^p}) }{|\boldsymbol{r_d^k}-\boldsymbol{r_d^p}|^3},
\label{eq:Fdd}
\end{equation}
where superscript $k$ denotes the dust particle, force on which is calculated. Summation is carried out over all simulated dust particle. 
\item The force, arising from the direct momentum transfer, when ions collide with dust particles
\begin{equation}
\boldsymbol{F_{id}^c} = \frac{m_i}{\delta t}\sum_{k} \boldsymbol{v_i^k},
\label{eq:Fidc}
\end{equation}
where $\boldsymbol{v_i^k}$ is the ion velocity, $m_i$ is the ion mass, $\delta t$ is the integration time step. Summation is carried out over all ions intersected the dust particle surface at the given time step. 
\item The force from the flow driving electric field, if given 
\begin{equation}
\boldsymbol{F_{ext}} = Q_d\boldsymbol{E_{ext}},
\label{eq:Fext}
\end{equation}
where $\boldsymbol{E_{ext}}$ is the external homogeneous electric field. 

 \end{enumerate}

 Charges of dust particles play an important role in the force calculation. There are three options for the charges calculation in OpenDust. The first one is the predetermined charges which are constants during a simulation. The second one is the calculation of dust particle charges from the orbital motion limited theory \cite{allen1992probe}. That option is suitable when one isolated dust particle is immersed to a weakly-collisional Maxwellian plasma flow. The third one is the most general one. It assumes the calculation of the electron charge flux via the orbital motion limited theory and the ion flux is calculated self-consistently. Each ion, intersecting the dust particle surface, is added to the ion charge flux. That option can be used in an arbitrary system of dust particles when electrons can be treated as Boltzmann fluid.

\section{Code capabilities and structure}

OpenDust aims to facilitate research in the field of complex plasma providing for researchers efficient and ease-of-use environment for simulation flowing plasma around dust grains and calculation forces, acting on these grains. It is designed for a wide range of users: from experimentalists to theoretical physicists, from students approaching complex plasma physics for the first time to seasoned researchers.

OpenDust is entirely written in Python and relies on the GPU-accelerated library for molecular dynamics OpenMM \cite{eastman2017openmm}. OpenMM is used for calculation of ions' trajectories, which is typically the "bottle neck" of simulation of flowing plasma around dust grains. Therefore, OpenDust inherits highly-optimized back-end with ability to work with GPU cards from OpenMM. In addition, OpenDust relies on the most common Python scientific packages, such as CuPy \cite{cupy_learningsys2017}, NumPy \cite{harris2020array}, SciPy \cite{2020SciPy-NMeth}, which all provide a solid foundation built, optimized, and well-documented by one of the largest community of developers.

OpenDust can be run by using a Python script or interactive environment such as Jupyter Notebook. Main simulation capabilities of OpenDust are briefly listed in Table 1 and described in the previous section.

\begin{figure}
    \centering
    \includegraphics[width=0.8\textwidth]{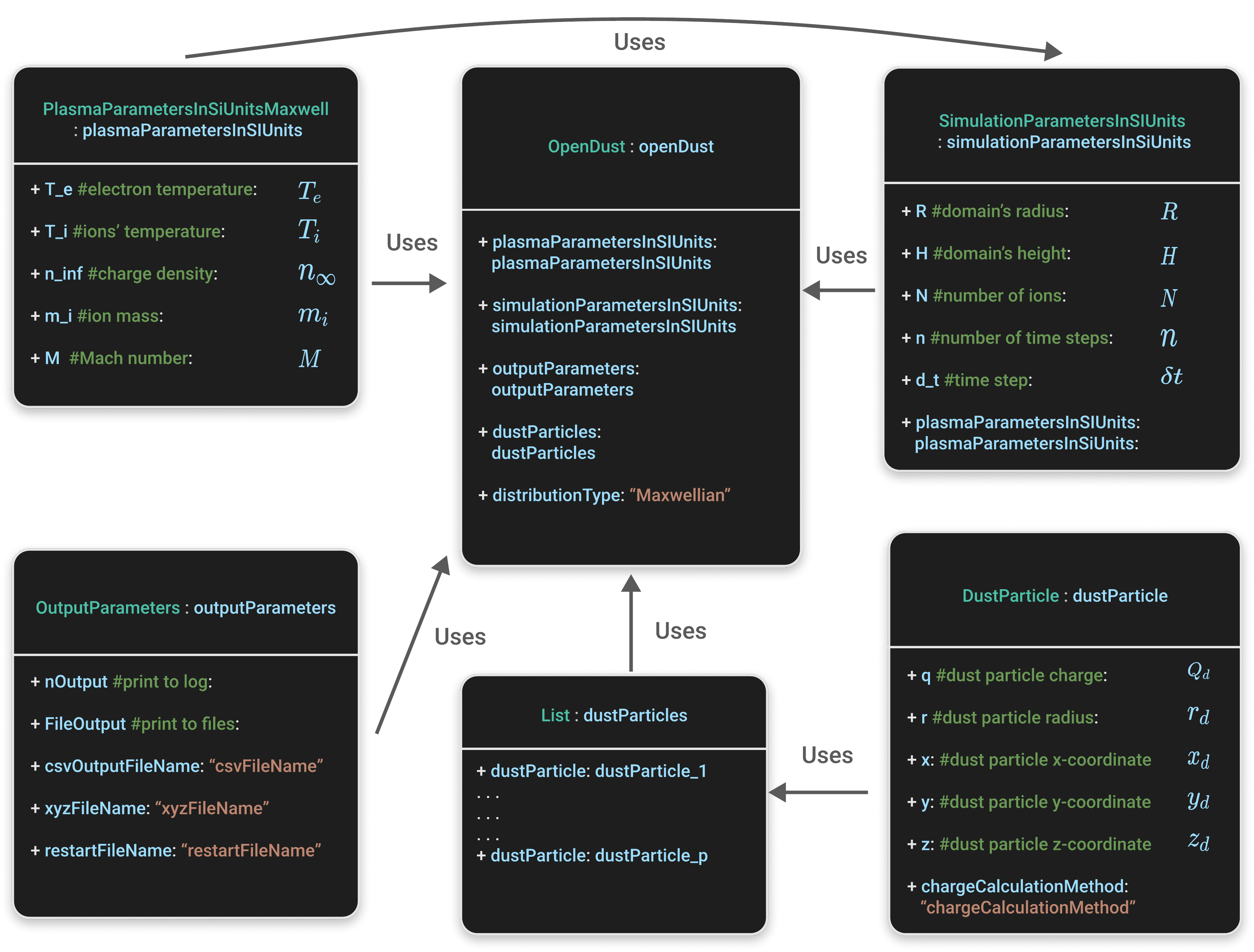}% 
    %\hspace{0.01 cm}
\caption{\label{fig:objectMaxwell} Interaction diagram of OpenDust class objects for the case of collisionless Maxwellian plasma flow.}
\end{figure}

\begin{figure}
    \centering
    \includegraphics[width=0.8\textwidth]{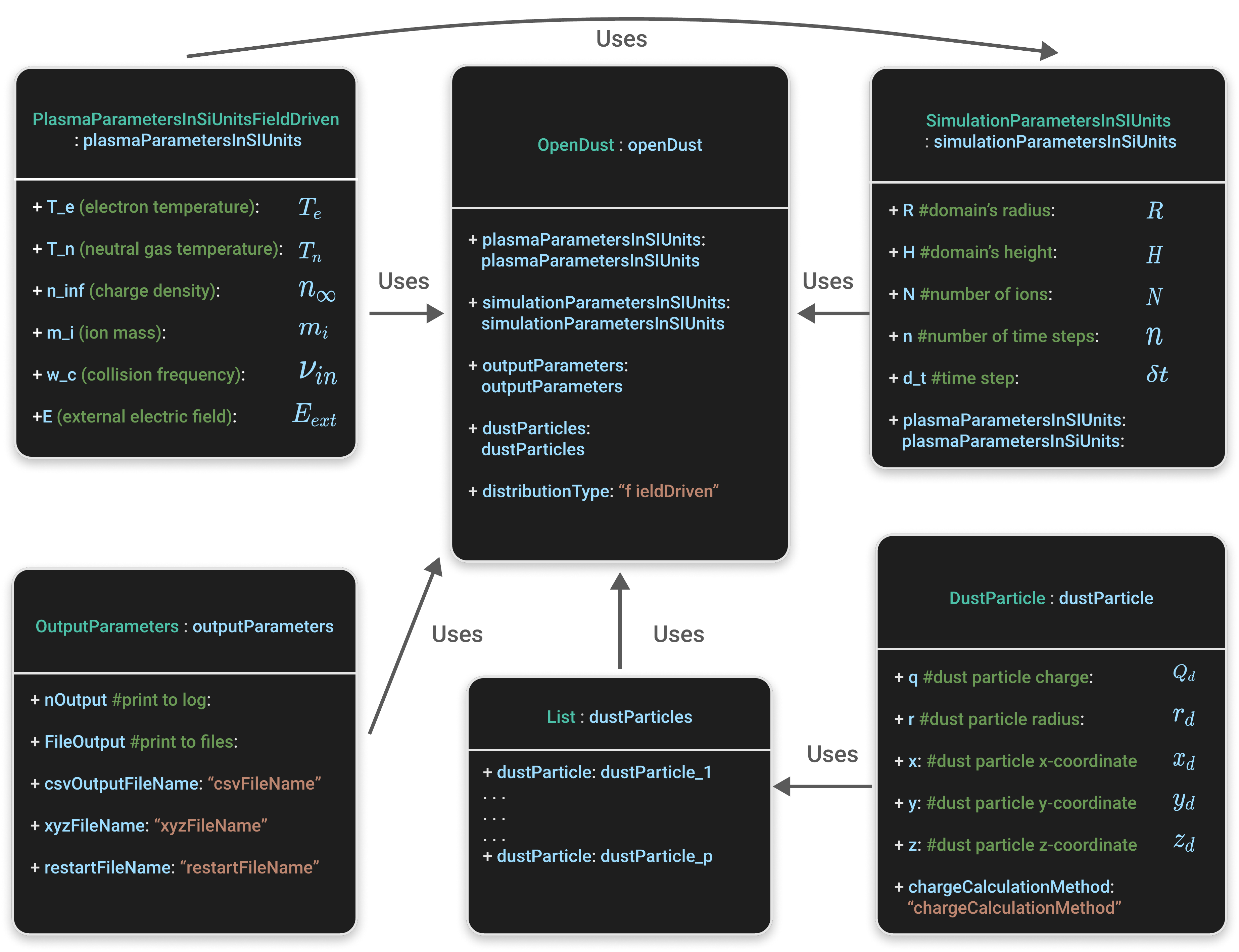}% 
    %\hspace{0.01 cm}
\caption{\label{fig:objectFieldDriven} Interaction diagram of OpenDust class objects for the case of field driven collisional plasma flow.}
\end{figure}

The main class of the package is \texttt{OpenDust}. This class stores the information about a simulated system and has the main method \texttt{simulate()} to perform calculations. In each simulation, \texttt{OpenDust} class object should be defined. Four auxiliary objects are needed to construct \texttt{OpenDust} class object. The interaction between main and the auxiliary objects is illustrated in the Figures \ref{fig:objectMaxwell} and \ref{fig:objectFieldDriven}. The auxiliary classes share simulation launching process in a logical manner: definition of plasma, simulation, output and dust particles parameters. OpenDust operates with parameters in SI units.

\texttt{Plasma\-Parameters\-In\-SI\-Units\-Maxwell} class or \texttt{Plasma\-Parameters\-In\-SI\-Units\-Field\-Driven} class are used to define plasma parameters in case of collisionless Maxwellian plasma flow or in case of field driven collisional plasma flow correspondingly. For both classes user should set value of electron temperature $T_e$, concentration of ions in the quasi-neutral region $n_{\infty}$, ion mass $m_i$. \texttt{Plasma\-Parameters\-In\-SI\-Units\-Maxwell} class requires additionally definition of Mach number of a plasma flow $M$ and \texttt{Plasma\-Parameters\-In\-SI\-Units\-Field\-Driven} class requires value of a driving electric field $E_{ext}$ and ion-neutral collision frequency $\nu_{in}$.

\texttt{Simulation\-Parameters\-In\-SI\-Units} class is used for definition of simulation parameters such as cylindrical domain radius $R$ and height $H$, number of discrete plasma particles (clouds of ions) $N$, number of integration time steps $n$, and value of an integration time step $\delta t$. As input parameter \texttt{Simulation\-Parameters\-In\-SI\-Units} class object takes also \texttt{Plasma\-Parameters\-In\-SI\-Units\-Maxwell} or \texttt{Plasma\-Parameters\-In\-SI\-Units\-Field\-Driven} class object.

OpenDust offers both console and file output during simulation. Parameters of the output can be defined via \texttt{Output\-Parameters} class. \texttt{Output\-Parameters} class constructor takes five positional arguments: \texttt{n\-Output}, \texttt{n\-File\-Output}, \texttt{csv\-Output\-File\-Name}, \texttt{xyz\-Output\-File\-Name}, \texttt{restart\-File\-Name}. In the simulation, some information is printed to a console every \texttt{n\-Output} integration time step and ion positions are written down to a file every \texttt{n\-File\-Output} integration time step. Default values for \texttt{n\-Output} and \texttt{n\-File\-Output} are ten. Ion positions can be written down in \texttt{.xyz} or \texttt{.csv} file formats. Data in these formats can be easily visualized, for example, using ParaView or Ovito open-source software. Absolute paths of \texttt{.xyz} and \texttt{.csv} files can be specified with \texttt{xyz\-Output\-File\-Name} and \texttt{csv\-Output\-File\-Name} correspondingly. User can also define the name of the file from which simulation can be restarted \texttt{restart\-File\-Name}. Default value of output file name is empty string, which means that data is not recorded.

Dust particles parameters are defined using Python list of \texttt{Dust\-Particle} class objects. If there are no dust particles in simulation Python list should be left empty. For every dust particle separate \texttt{Dust\-Particle} class object must be created. An initial charge $Q_d$, radius $r_d$, and three position coordinates $x_d$, $y_d$, $z_d$  of the dust particle should be defined to create \texttt{Dust\-Particle} class object. The method of calculation dust particles' charges can be defined via  \texttt{charge\-Calculation\-Method} parameter. Three options are available:  constant charge \texttt{"given"}, orbital motion limited charge calculation \texttt{"oml"} and self-consistent charge calculation \texttt{"consistent"} as described in Section 2. Charge calculation method must be the same for all \texttt{Dust\-Particle} class objects using in simulation.

When all four auxiliary objects are defined the \texttt{Open\-Dust} class object can be constructed. Constructor of \texttt{Open\-Dust} class requires one more necessary argument \texttt{distribution\-Type}, which describes whether the case of collisionless Maxwellian plasma flow \texttt{"Maxwellian"} or the case of field driven collisional plasma flow \texttt{"field\-Driven"} is utilized. Note that \texttt{"Maxwellian"} only can be used with  \texttt{Plasma\-Parameters\-In\-SI\-Units\-Maxwell} class object and \texttt{"field\-Driven"} only with \texttt{Plasma\-Parameters\-In\-SI\-Units\-Field\-Driven} class object.

Simulation can be launched using \texttt{simulate()} method of an \texttt{Open\-Dust} class object. \texttt{simulate()} takes three optional keyword arguments \texttt{device\-Index}, \texttt{cut\-Off}, \texttt{to\-Restart\-File\-Name}. Via the  \texttt{device\-Index} argument, indexes of GPU devices needed for calculations can be set. \texttt{device\-Index} argument value is a string of comma separated device indexes with the default \texttt{"0"} string, i.e. only one GPU device is used.
\texttt{cut\-Off} argument can be \texttt{True} or \texttt{False} with the \texttt{False} default. \texttt{cut\-Off} argument describes whether use cutoff distance for ion-ion interactions calculation or not.
The default value of the \texttt{to\-Restart\-File\-Name} argument is an empty string. If the argument is not an empty string then OpenDust will try to restart simulation using restart file  \texttt{to\-Restart\-File\-Name}.

\begin{figure}
    \centering
    \includegraphics[width=0.8\textwidth]{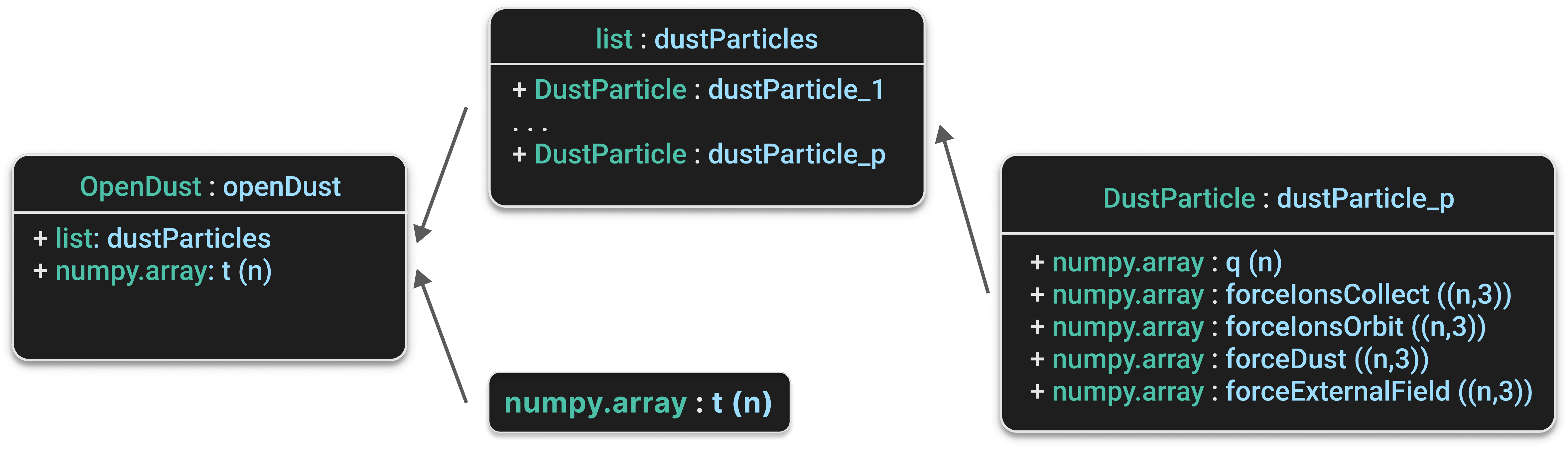}% 
    %\hspace{0.01 cm}
\caption{\label{fig:objectResult} dependence diagram for \texttt{t} and \texttt{dust\-Particles} attributes.}
\end{figure}

As \texttt{Open\-Dust} class object is initialized its attributes \texttt{t} and \texttt{dust\-Particles} are created. When simulation of a plasma flow is completed, these attributes can be used to analyze time-dependent charges of dust particles and forces, acting on them. dependence diagram for these attributes is presented in Figure \ref{fig:objectResult}. \texttt{t(n)} is the one-dimensional \texttt{numpy} array, which stores discrete time points where system state is calculated. The length of the array is the number of integration time steps \texttt{n}. The units of time steps are seconds. \texttt{dust\-Particles} is the Python list, each element of that is an \texttt{Open\-Dust.\-Dust\-Particle} class object associated with a separate dust particle. \texttt{Open\-Dust.\-Dust\-Particle} class object has five attributes: charge of a dust particle \texttt{q(n)}, force from electric interactions between ions and a dust particle \texttt{force\-Ions\-Orbit((n,3))} \eqref{eq:Fido}, dust-dust electric interaction force \texttt{force\-Dust((n,3))} \eqref{eq:Fdd}, force from momentum transfer via direct collisions between ions and a dust particle  \texttt{forceIonsCollect((n,3))} \eqref{eq:Fidc}, force from external homogeneous electric field \texttt{force\-External\-Field((n,3))} \eqref{eq:Fext}. \texttt{q(n)} is the one-dimensional \texttt{numpy} array and \texttt{force\-Ions\-Orbit((n,3))}, \texttt{force\-Dust((n,3))}, \texttt{force\-Ions\-Collect((n,3))}, \texttt{force\-External\-Field((n,3))} are two-dimensional \texttt{numpy} arrays. The first dimension of the attribute arrays stores time series of charge and force. The second dimension of the force arrays is used to store three space component of the forces. 

\section{Test cases}

In order to illustrate the capabilities of OpenDust we present several test-cases. These tests can be found in the \texttt{example} directory of the OpenDust repository. In addition, the tests are intended to verify  the code against previous calculations of complex plasma.

\subsection{Solitary dust particle in a collisionless Maxwellian plasma flow}
\begin{figure}
    \centering
    \includegraphics[width=1\textwidth]{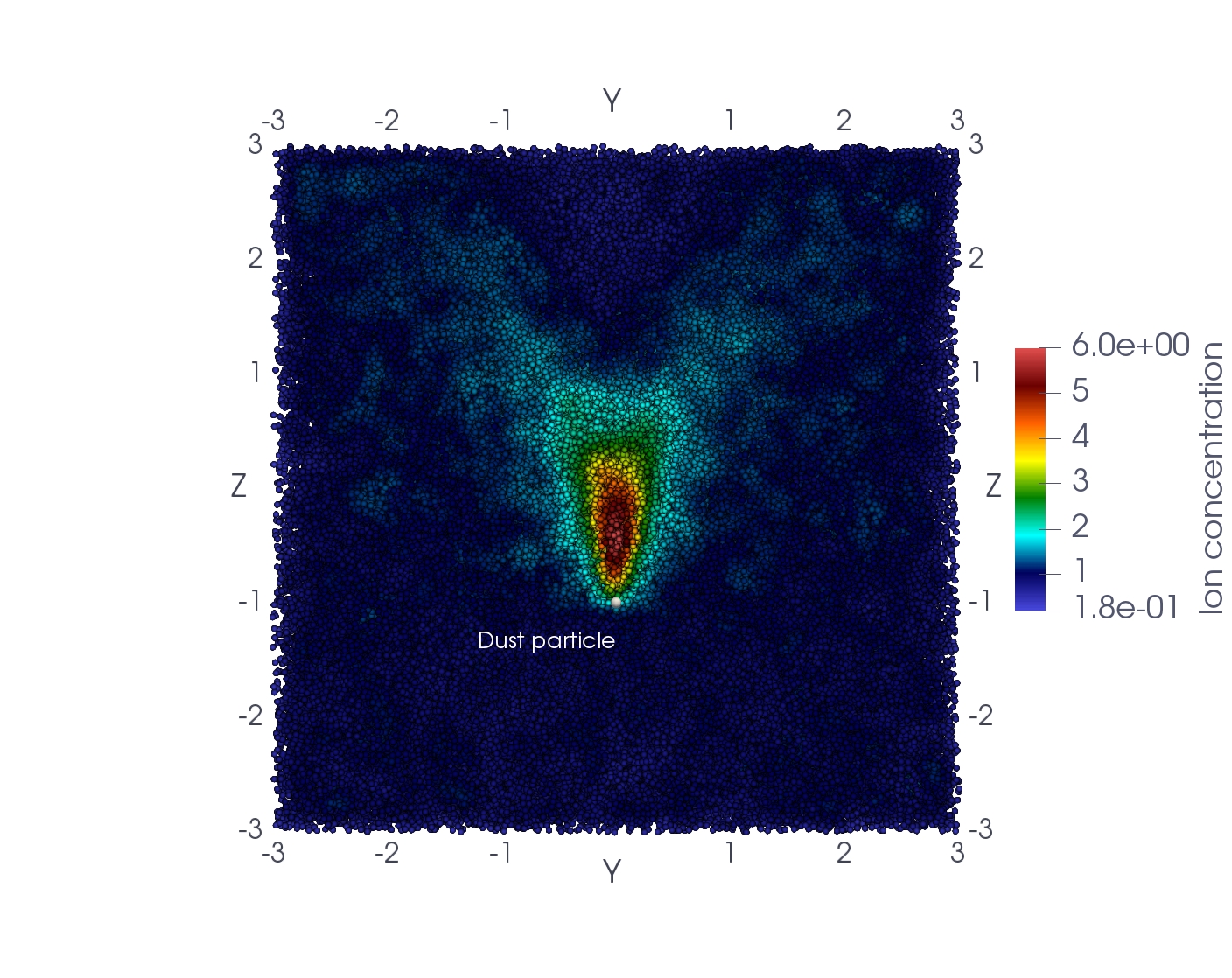}% 
    %\hspace{0.01 cm}
\caption{\label{fig:collisionlessAnimation} Stationary ion density distribution around a solitary dust particle in a plasma flow.  Coordinates are given in electron Debye radius units and ion concentration is normalized on ion concentration in the unperturbed plasma area. Ion flow is directed to the top. Video of the ion density distribution changing during the simulation is available online. Multimedia view: }
\end{figure}

A solitary dust particle in a collisionless Maxwellian plasma flow is considered. Firstly, we present simulation results for a single value of plasma flow velocity. Used parameters are presented in Table 2. Dust particle charge is calculated using all three available in OpenDust options: preset charge, orbital motion limited charge calculation (OML), self-consistent charge calculation. The first simulation is carried out with the preset value of the dust particle charge $Q_d = 392500$ electron charges. View of the calculated stationary ion density distribution is presented in the Figure \ref{fig:collisionlessAnimation}. The dust particle is pictured as a white ball and ion flow is directed to the top. Ion density perturbation takes characteristic cone-shaped structure in accordance with \cite{miloch2010wake} and has positive maximum behind the dust particle.

\begin{figure}
    \centering
    \includegraphics[width=0.8\textwidth]{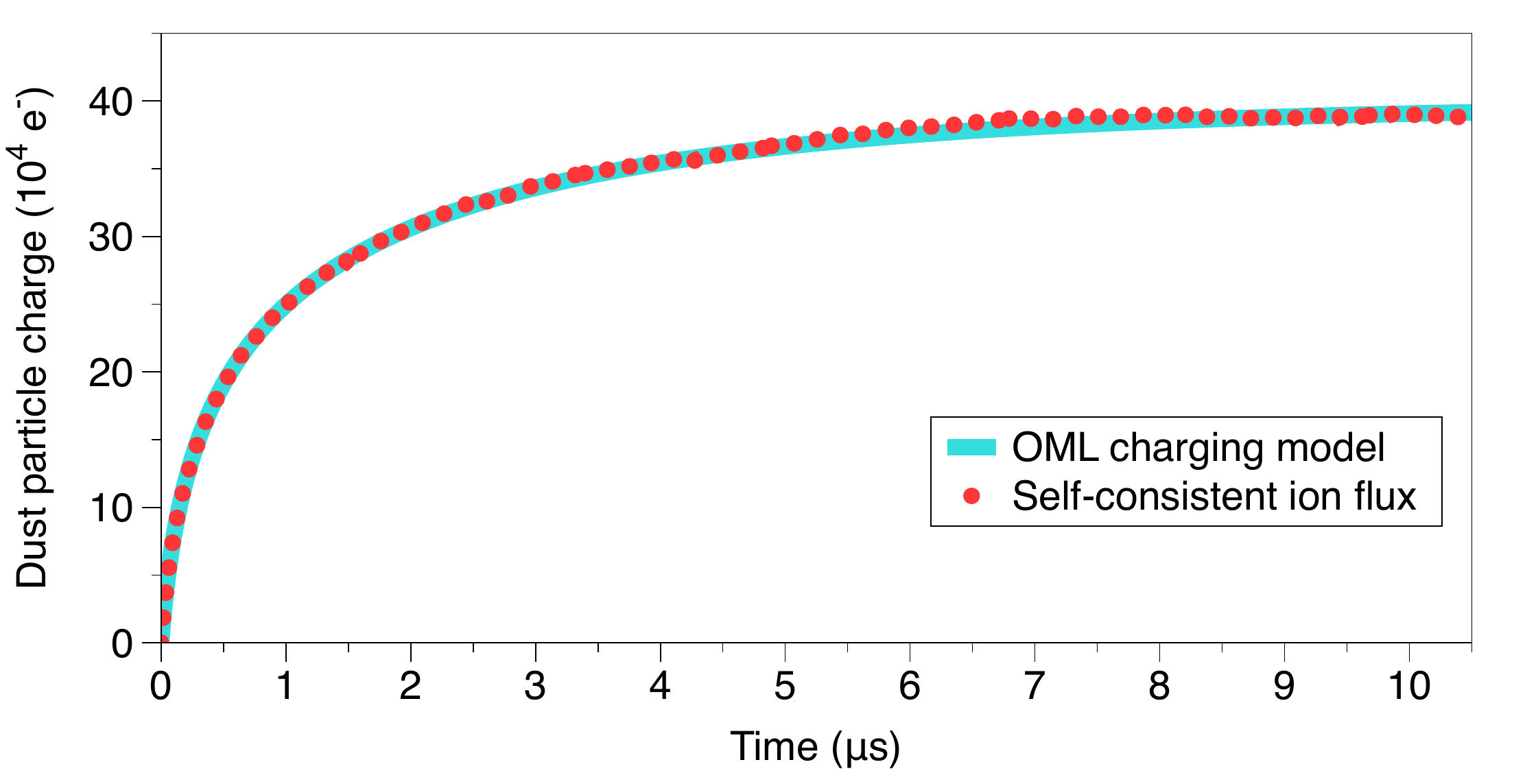}% 
\caption{\label{fig:collisionlessCharge} Comparison of a time-dependent dust particle charge calculated self-consistently and using OML approximation. }
\end{figure}

In collisionless Maxwellian plasma flow, charging process of a solitary dust particle can be described using OML approximation \cite{matthews2020dust}. In order to test correctness of dust particle charge calculation in OpenDust we compare time-dependent dust particle charge calculated self-consistently and via OML. The comparison is presented in the Figure \ref{fig:collisionlessCharge}. As one can see, two curves match well each other.  

\begin{table}[ht!]
  \begin{center}
    \caption{Parameters used for simulation of collisionless Maxwellian plasma flow around a solitary dust particle for test-case number one}
    \label{tab:table2}
    \begin{tabular}{|p{0.5\textwidth}|p{0.2\textwidth}|p{0.2\textwidth}|}
      \hline
      \raggedright Parameter & \raggedright Value & \raggedright Units \tabularnewline 
      \hline
    \end{tabular}
    \begin{tabular}{p{0.9\textwidth}}
      \centering Plasma parameters \tabularnewline 
    \end{tabular}
    \begin{tabular}{|p{0.5\textwidth}|p{0.2\textwidth}|p{0.2\textwidth}|}
      \hline
      \raggedright Electron temperature $T_e$ & \raggedright $29011$ & \raggedright K \tabularnewline 
      \hline
      \raggedright Ion temperature $T_i$ & \raggedright $290.11$ & \raggedright K \tabularnewline
      \hline
      \raggedright Concentration of charged plasma species in the quasi-neutral region $n_{\inf}$ & \raggedright $1e14$ & \raggedright $1/m^3$ \tabularnewline
      \hline
      \raggedright Ion mass $m_i$ & \raggedright $ 6.6335209\mathrm{e}{-26}$ & \raggedright kg \tabularnewline
      \hline
      \raggedright Mach number of a plasma flow $M$ & \raggedright $ 1$ & \raggedright - \tabularnewline
      \hline
    \end{tabular}
    \begin{tabular}{p{0.9\textwidth}}
      \centering Simulation parameters \tabularnewline 
    \end{tabular}
    \begin{tabular}{|p{0.5\textwidth}|p{0.2\textwidth}|p{0.2\textwidth}|}
      \hline
      \raggedright Radius of the cylindrical computational domain $R$ & \raggedright $3$ & \raggedright electron Debye radius \tabularnewline 
      \hline
      \raggedright Height of the cylindrical computational domain $H$ & \raggedright $6$ & \raggedright electron Debye radius \tabularnewline
      \hline
      \raggedright Number of discrete particles $N$ & \raggedright $2^{16}$ & \raggedright - \tabularnewline
      \hline
      \raggedright Number of integration time steps $n$ & \raggedright $3000$ & \raggedright - \tabularnewline
      \hline
      \raggedright Integration time step & \raggedright $3.5148\mathrm{e}{-10}$ & \raggedright s \tabularnewline
      \hline
    \end{tabular}
    \begin{tabular}{p{0.9\textwidth}}
      \centering Dust particle parameters \tabularnewline 
    \end{tabular}
    \begin{tabular}{|p{0.5\textwidth}|p{0.2\textwidth}|p{0.2\textwidth}|}
      \hline
      \raggedright Dust particle radius $r_d$ & \raggedright $58.8\mathrm{e}{-6}$ & \raggedright m \tabularnewline 
      \hline
      \raggedright Dust particle position  $(x_d;y_d;z_d)$ & \raggedright $(0;0;-1)$ & \raggedright electron Debye radius \tabularnewline 
      \hline
    \end{tabular}
  \end{center}
 \end{table}

\begin{figure}
    \centering
    \includegraphics[width=1\textwidth]{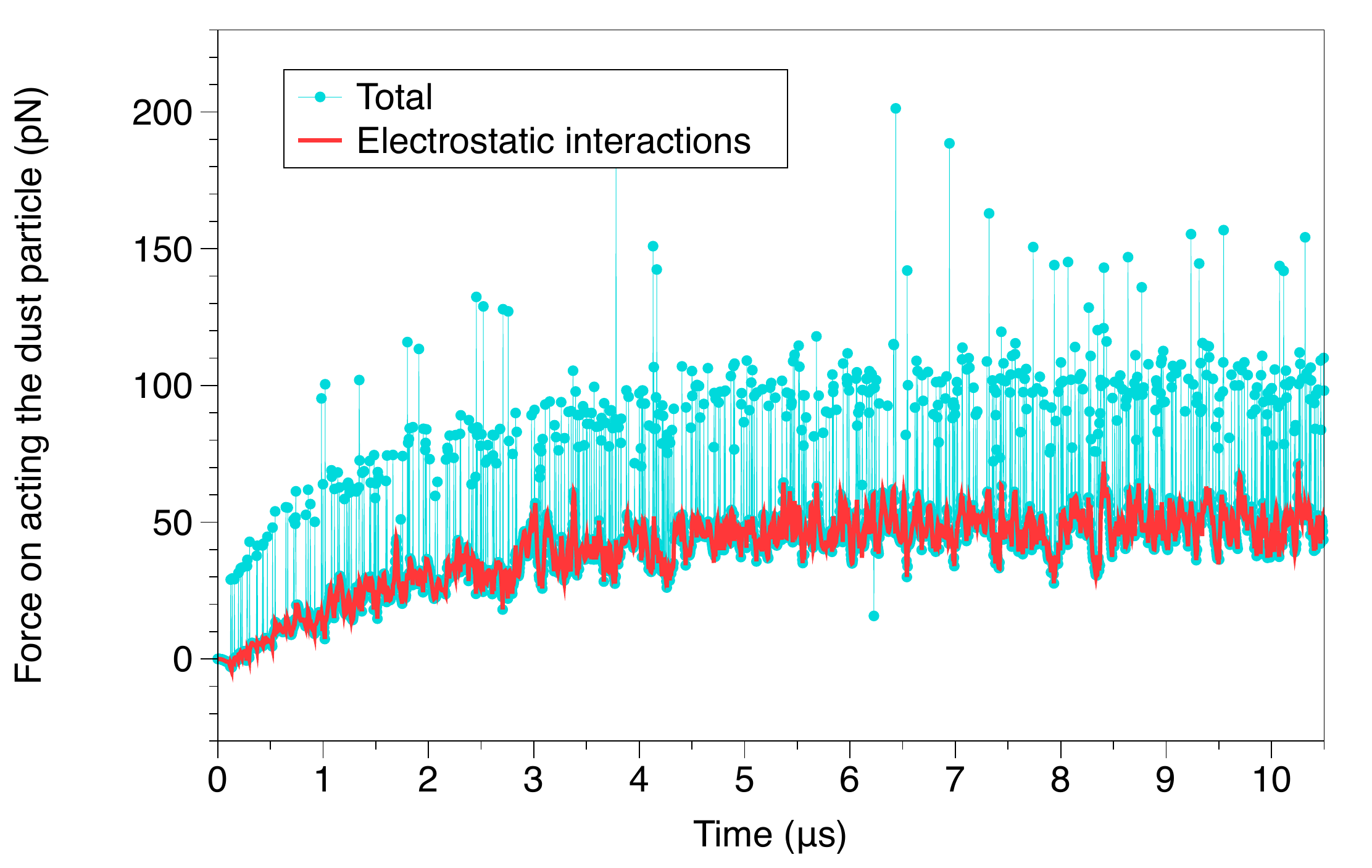}
\caption{\label{fig:collisionlessForce} Time dependence of the ion drag force. The light blue line shows total ion drag force and the red line shows force term from ion-dust electric interactions.}
\end{figure}

In addition, we calculate total force, acting on the solitary dust from the plasma flow, so-called ion drag force. As explained in the Section 2, calculated ion drag force in OpenDust consists of two parts: the electric force between dust and shielded ions \eqref{eq:Fido} and the force, arising from the direct momentum transfer in ion-dust collisions \eqref{eq:Fidc}. In the Figure \ref{fig:collisionlessForce}, time dependences of total force and electric force term are presented. Total force has greater fluctuations due to the momentum transfer force term. After approximately eight microseconds, the total force and the dust particle charge reach stationary value.

Computational time needed for such simulation is varied depending on the used GPU-devices from several seconds to two minutes. In comparison, Particle-In-Cell CPU-based code Coptic \cite{hutchinson2011nonlinear} spends about an hour.

For OpenDust verification we calculate dependence of the ion drag force stationary value on ion flow velocity. This dependence is compared against two previous calculations of the ion drag force \cite{hutchinson2006collisionless, piel2017molecular}. Results from \cite{hutchinson2006collisionless} are calculated using Particle-In-Cell method. In \cite{piel2017molecular} ion drag force is calculated according to Piel's asymmetric molecular dynamics. The plasma  parameters for all three curves are the same and listed in the Table 2. The dust particle charge is fixed during simulation and corresponds to the OML stationary value for a given flow velocity. In the reference articles collisions between dust particle and ions are not considered, so the mean Opendust force is calculated from the electric interactions between dust and shielded ions. Ion drag force calculated with OpenDust matches well  asymmetric molecular dynamics and Particle-In-Cell results.

\begin{figure}
    \centering
    \includegraphics[width=0.8\textwidth]{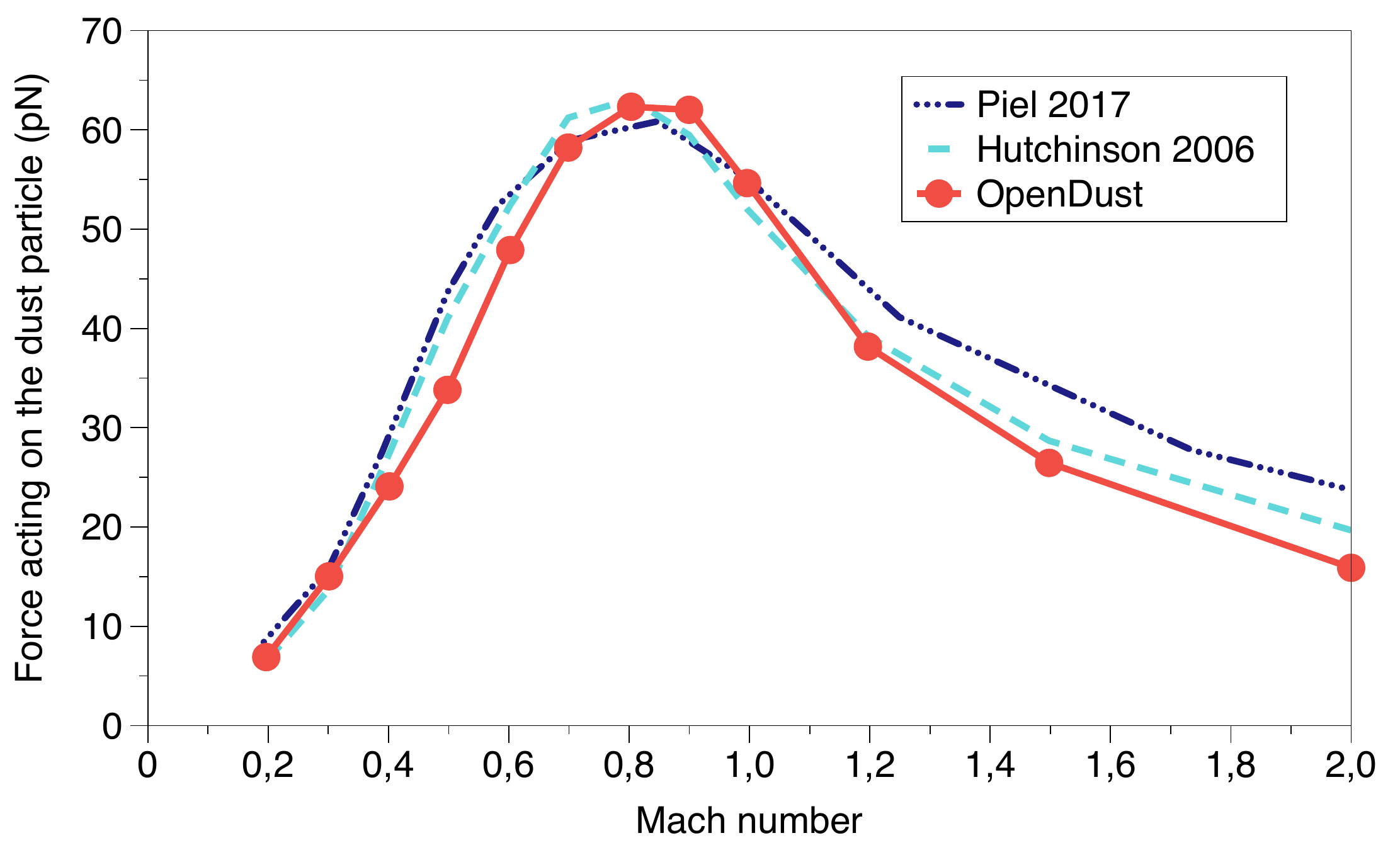}% 
\caption{\label{Mach} Comparison of the ion drag force with the results from previous calculations \cite{hutchinson2006collisionless, piel2017molecular}. OpenDust results are depicted with red points with the lines serving to guide the eye. The light blue dashed line and the dark blue dashed-dotted line show results from references \cite{hutchinson2006collisionless} and \cite{piel2017molecular} correspondingly.}
\end{figure}

\subsection{Solitary dust particle in a field driven collision
flow}

\begin{figure}
    \centering
    \includegraphics[width=1\textwidth]{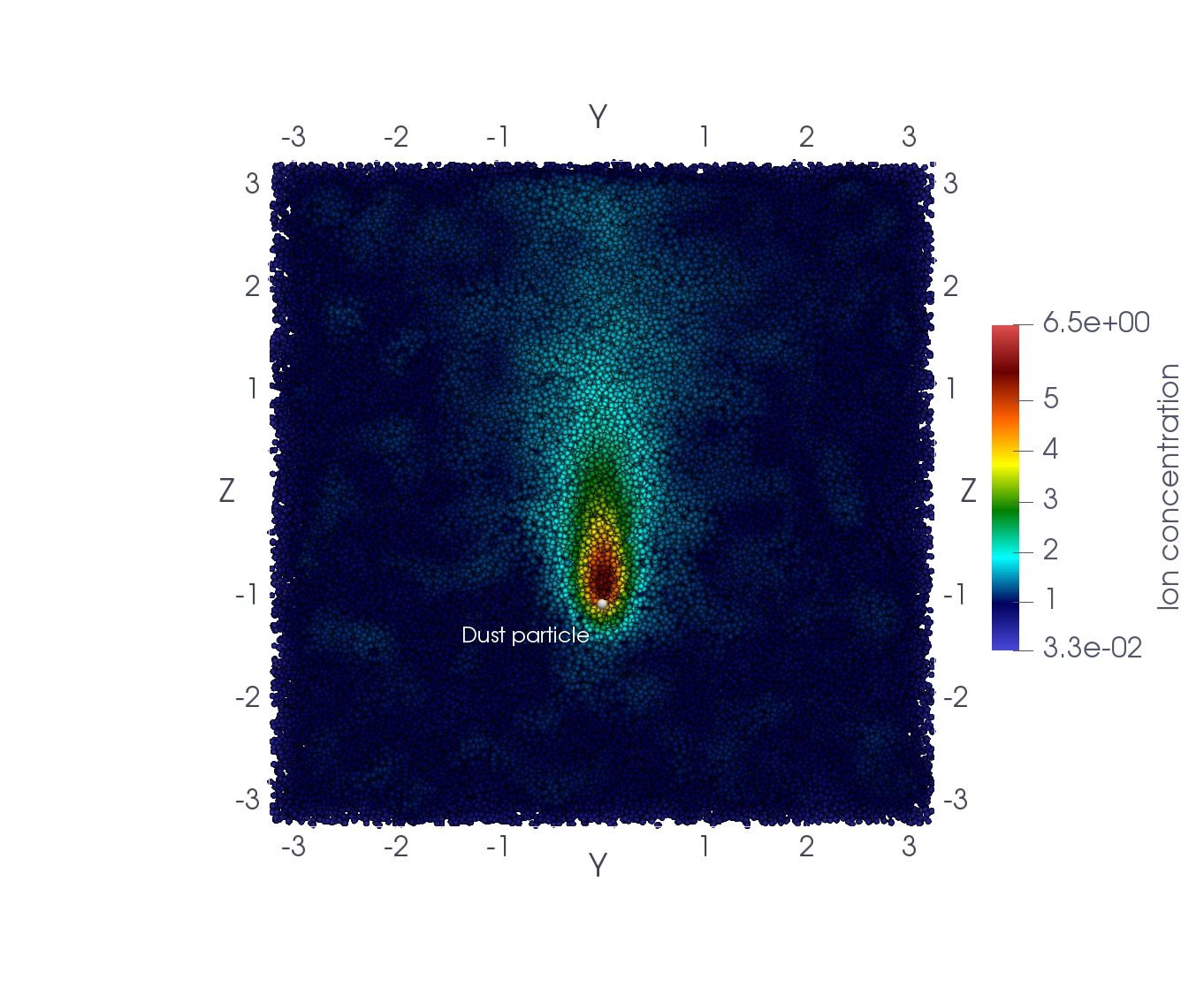}% 
    %\hspace{0.01 cm}
\caption{\label{fig:collisionalAnimation} Stationary ion density distribution around a solitary dust particle in a field driven collisional plasma flow. Coordinates are given in electron Debye radius units and ion concentration is normalized on ion concentration in the unperturbed plasma area. Ion flow is directed to the top. Video of the ion density distribution changing during the simulation is available online. Multimedia view: }
\end{figure}

In this section, simulation results of a solitary dust particle embedded in a field driven collisional plasma flow is presented. Typical view of calculated ion density distribution around a dust particle in a collisional plasma flow is presented in the Figure \ref{fig:collisionalAnimation}. In contrast to the collisionless case, ion density perturbation behind a dust particle does not have the cone structure and more suppressed. These corresponds to the conclusions of the previous studies \cite{sundar2017impact}.

Particle-In-Cell simulation results from \cite{hutchinson2007computation} are used here as a reference for the OpenDust simulations. Stationary values of ion drag force and self-consistent dust particle charge  for different ion-neutral collision frequency are calculated. Driving electric field is selected so that the ion flow velocity is equal to one Bohm velocity for each value of collision frequency.  Used parameters are presented in Table \ref{tab:table3}. Integration time step is selected to be at least five times smaller than characteristic ion-neutral collision time.

\begin{table}[ht!]
  \begin{center}
    \caption{Parameters used for simulation of field driven collisional plasma flow around a solitary dust particle}
    \label{tab:table3}
    \begin{tabular}{|p{0.5\textwidth}|p{0.2\textwidth}|p{0.2\textwidth}|}
      \hline
      \raggedright Parameter & \raggedright Value & \raggedright Units \tabularnewline 
      \hline
    \end{tabular}
    \begin{tabular}{p{0.9\textwidth}}
      \centering Plasma parameters \tabularnewline 
    \end{tabular}
    \begin{tabular}{|p{0.5\textwidth}|p{0.2\textwidth}|p{0.2\textwidth}|}
      \hline
      \raggedright Electron temperature $T_e$ & \raggedright $30000$ & \raggedright K \tabularnewline 
      \hline
      \raggedright Ion temperature $T_i$ & \raggedright $300$ & \raggedright K \tabularnewline
      \hline
      \raggedright Concentration of charged plasma species in the quasi-neutral region $n_{\inf}$ & \raggedright $3.57\mathrm{e}{15}$ & \raggedright $1/m^3$ \tabularnewline
      \hline
      \raggedright Ion mass $m_i$ & \raggedright $ 1.673557\mathrm{e}{-27}$ & \raggedright kg \tabularnewline
      \hline
      \raggedright External electric field $E_{ext}$ & \raggedright $ 8.117$\\...\\$2.647{e}{6}$ & \raggedright V/m \tabularnewline
      \hline
      \raggedright Ion-neutral collision frequency $E_{ext}$ & \raggedright $ 4.94{e}{6}$\\...\\$1.61{e}{10}$ & \raggedright Hz \tabularnewline
      \hline
    \end{tabular}
    \begin{tabular}{p{0.9\textwidth}}
      \centering Simulation parameters \tabularnewline 
    \end{tabular}
    \begin{tabular}{|p{0.5\textwidth}|p{0.2\textwidth}|p{0.2\textwidth}|}
      \hline
      \raggedright Radius of the cylindrical computational domain $R$ & \raggedright $1.25$ & \raggedright electron Debye radius \tabularnewline 
      \hline
      \raggedright Height of the cylindrical computational domain $H$ & \raggedright $6$ & \raggedright electron Debye radius \tabularnewline
      \hline
      \raggedright Number of discrete particles $N$ & \raggedright $2^{17}$ & \raggedright - \tabularnewline
      \hline
      \raggedright Number of integration time steps $n$ & \raggedright $200000$ & \raggedright - \tabularnewline
      \hline
      \raggedright Integration time step  $\delta t$ & \raggedright $1\mathrm{e}{-11}$ & \raggedright s \tabularnewline
      \hline
    \end{tabular}
    \begin{tabular}{p{0.9\textwidth}}
      \centering Dust particle parameters \tabularnewline 
    \end{tabular}
    \begin{tabular}{|p{0.5\textwidth}|p{0.2\textwidth}|p{0.2\textwidth}|}
      \hline
      \raggedright Dust particle radius $r_d$ & \raggedright $1\mathrm{e}{-5}$ & \raggedright m \tabularnewline 
      \hline
      \raggedright Dust particle position  $(x_d;y_d;z_d)$ & \raggedright $(0;0;0)$ & \raggedright electron Debye radius \tabularnewline 
      \hline
    \end{tabular}
  \end{center}
 \end{table}

\begin{figure}
  \begin{subfigure}{0.5\textwidth}
    \includegraphics[width=\linewidth]{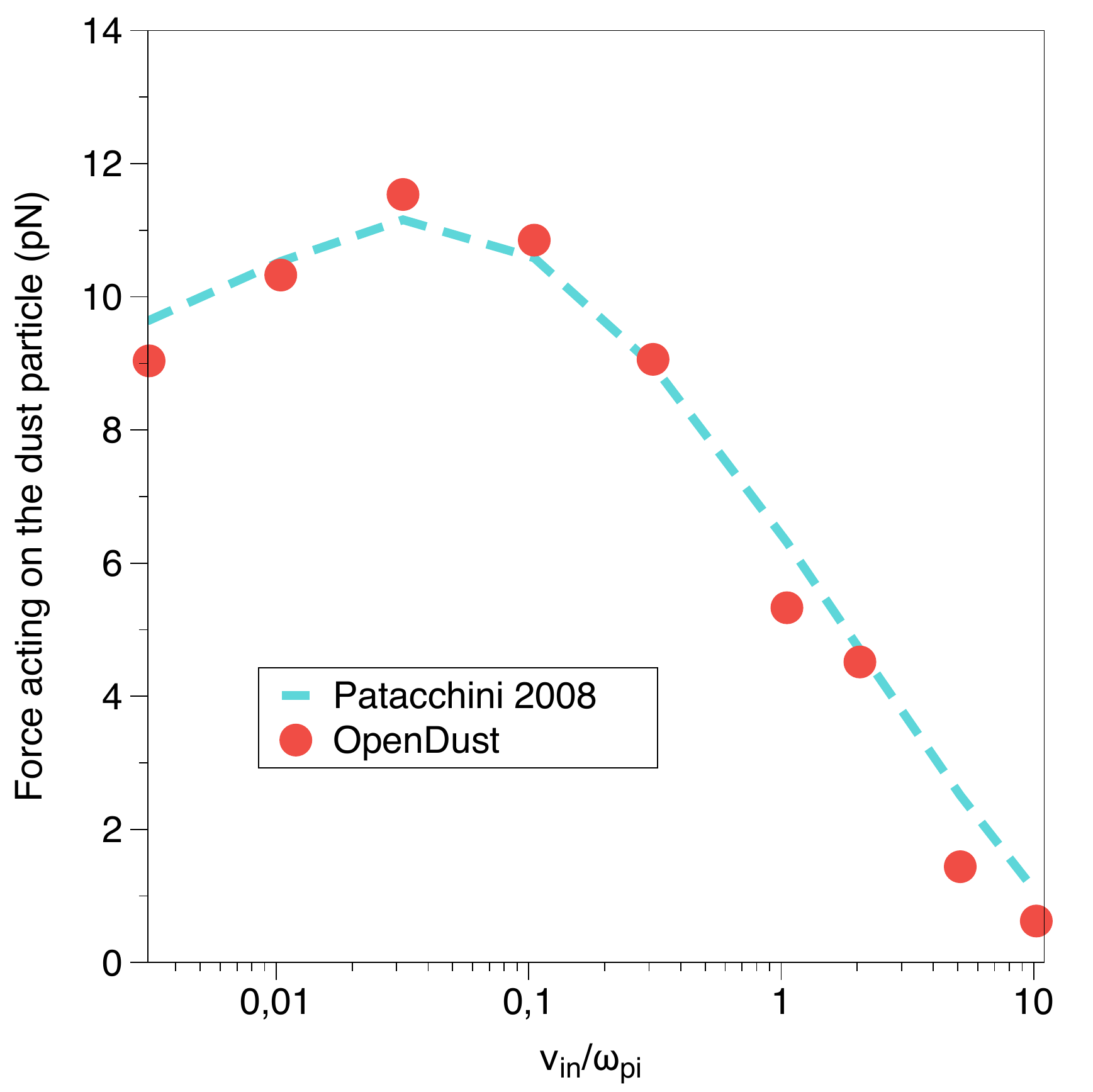}
    \caption{\label{fig:PatacchiniForce} Ion drag force}
  \end{subfigure}%
  \hspace*{\fill} 
  \begin{subfigure}{0.5\textwidth}
    \includegraphics[width=\linewidth]{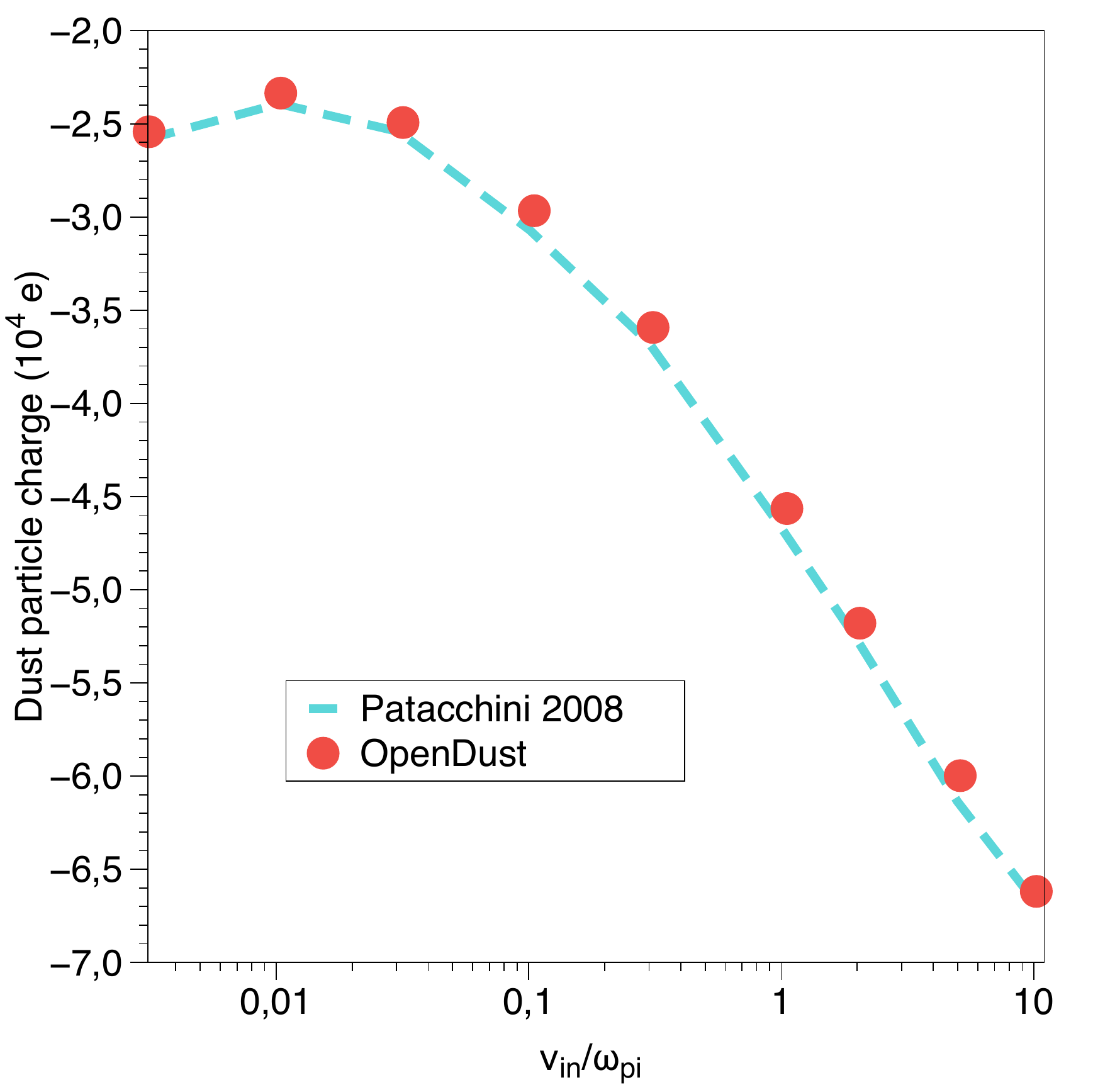}
    \caption{\label{fig:PatacchiniCharge} Dust particle charge}
  \end{subfigure}%
  \hspace*{\fill}
\caption{\label{fig:PatacchiniComparison} dependence of the ion drag force (a) and dust particle charge (b) on ion-neutral collision frequency. OpenDust results are depicted with blue points with the lines serving to guide the eye. The red dashed line shows results from \cite{hutchinson2007computation}.}
\end{figure}

Comparison between OpenDust and \cite{hutchinson2007computation} force and charge dependence is presented in Figure \ref{fig:PatacchiniComparison}. OpenDust results demonstrate good agreement with Particle-In-Cell calculation for both ion drag force and dust particle charge.

\subsection{Two dust particles in a collisionless Maxwellian plasma flow}

Above, we discussed the simulations of a plasma flow around a solitary dust particle. Here, the case of two interacted dust particles is described. Technically, launching script for the simulation of one dust particle and a group of dust particles is the same. When one dust particle is simulated \texttt{dust\-Particles} list contains only one  \texttt{Dust\-Particle} class object. Additional dust particles can be simply added to that list.

In order to check if the OpenDust simulates a group of dust particles properly, we compare forces, acting in a two-particle system with
the upper particle fixed at the position $x = 0, z = -0.5 r_{De}$ and the lower particle has a vertical and horizontal distance from
the upper particle $\Delta z = 1.0 r_{De}$ and $\Delta x = 0.5 r_{De}$ correspondingly, against the results of \cite{piel2017molecular}. Parameters are listed in the Table \ref{tab:table4}. Charges of dust particles are preset and plasma flow is collisionless and Maxwellian. In case of simulation a group of dust particles, dust-dust interactions \eqref{eq:Fdd} should be added to ion-dust forces in order to calculate the total force. 

\begin{table}[ht!]
  \begin{center}
    \caption{Parameters used for simulation of collisionless Maxwellian plasma flow around two dust particles}
    \label{tab:table4}
    \begin{tabular}{|p{0.5\textwidth}|p{0.2\textwidth}|p{0.2\textwidth}|}
      \hline
      \raggedright Parameter & \raggedright Value & \raggedright Units \tabularnewline 
      \hline
    \end{tabular}
    \begin{tabular}{p{0.9\textwidth}}
      \centering Plasma parameters \tabularnewline 
    \end{tabular}
    \begin{tabular}{|p{0.5\textwidth}|p{0.2\textwidth}|p{0.2\textwidth}|}
      \hline
      \raggedright Electron temperature $T_e$ & \raggedright $29011$ & \raggedright K \tabularnewline 
      \hline
      \raggedright Ion temperature $T_i$ & \raggedright $290.11$ & \raggedright K \tabularnewline
      \hline
      \raggedright Concentration of charged plasma species in the quasi-neutral region $n_{\inf}$ & \raggedright $1e14$ & \raggedright $1/m^3$ \tabularnewline
      \hline
      \raggedright Ion mass $m_i$ & \raggedright $ 6.6335209\mathrm{e}{-27}$ & \raggedright kg \tabularnewline
      \hline
      \raggedright Mach number of a plasma flow $M$ & \raggedright $ 0.3$\\...\\$1.0$ & \raggedright - \tabularnewline
      \hline
    \end{tabular}
    \begin{tabular}{p{0.9\textwidth}}
      \centering Simulation parameters \tabularnewline 
    \end{tabular}
    \begin{tabular}{|p{0.5\textwidth}|p{0.2\textwidth}|p{0.2\textwidth}|}
      \hline
      \raggedright Radius of the cylindrical computational domain $R$ & \raggedright $3$ & \raggedright electron Debye radius \tabularnewline 
      \hline
      \raggedright Height of the cylindrical computational domain $H$ & \raggedright $6$ & \raggedright electron Debye radius \tabularnewline
      \hline
      \raggedright Number of discrete particles $N$ & \raggedright $2^{17}$ & \raggedright - \tabularnewline
      \hline
      \raggedright Number of integration time steps $n$ & \raggedright $50000$ & \raggedright - \tabularnewline
      \hline
      \raggedright Integration time step & \raggedright $1\mathrm{e}{-9}$ & \raggedright s \tabularnewline
      \hline
    \end{tabular}
    \begin{tabular}{p{0.9\textwidth}}
      \centering Dust particles parameters \tabularnewline 
    \end{tabular}
    \begin{tabular}{|p{0.5\textwidth}|p{0.2\textwidth}|p{0.2\textwidth}|}
      \hline
      \raggedright Dust particle radii $r_d$ & \raggedright $12\mathrm{e}{-6}$ & \raggedright m \tabularnewline 
      \hline
      \raggedright Dust particle charges  $Q_d$ & \raggedright $40900$ & \raggedright electron charges \tabularnewline 
      \hline
      \raggedright First dust particle position  $(x_d;y_d;z_d)$ & \raggedright $(0;0;-0.5)$ & \raggedright electron Debye radius \tabularnewline 
      \hline
      \raggedright Second dust particle position  $(x_d;y_d;z_d)$ & \raggedright $(0.5;0;0.5)$ & \raggedright electron Debye radius \tabularnewline 
      \hline
    \end{tabular}
  \end{center}
 \end{table}

\begin{figure}
    \centering    \includegraphics[width=0.8\textwidth]{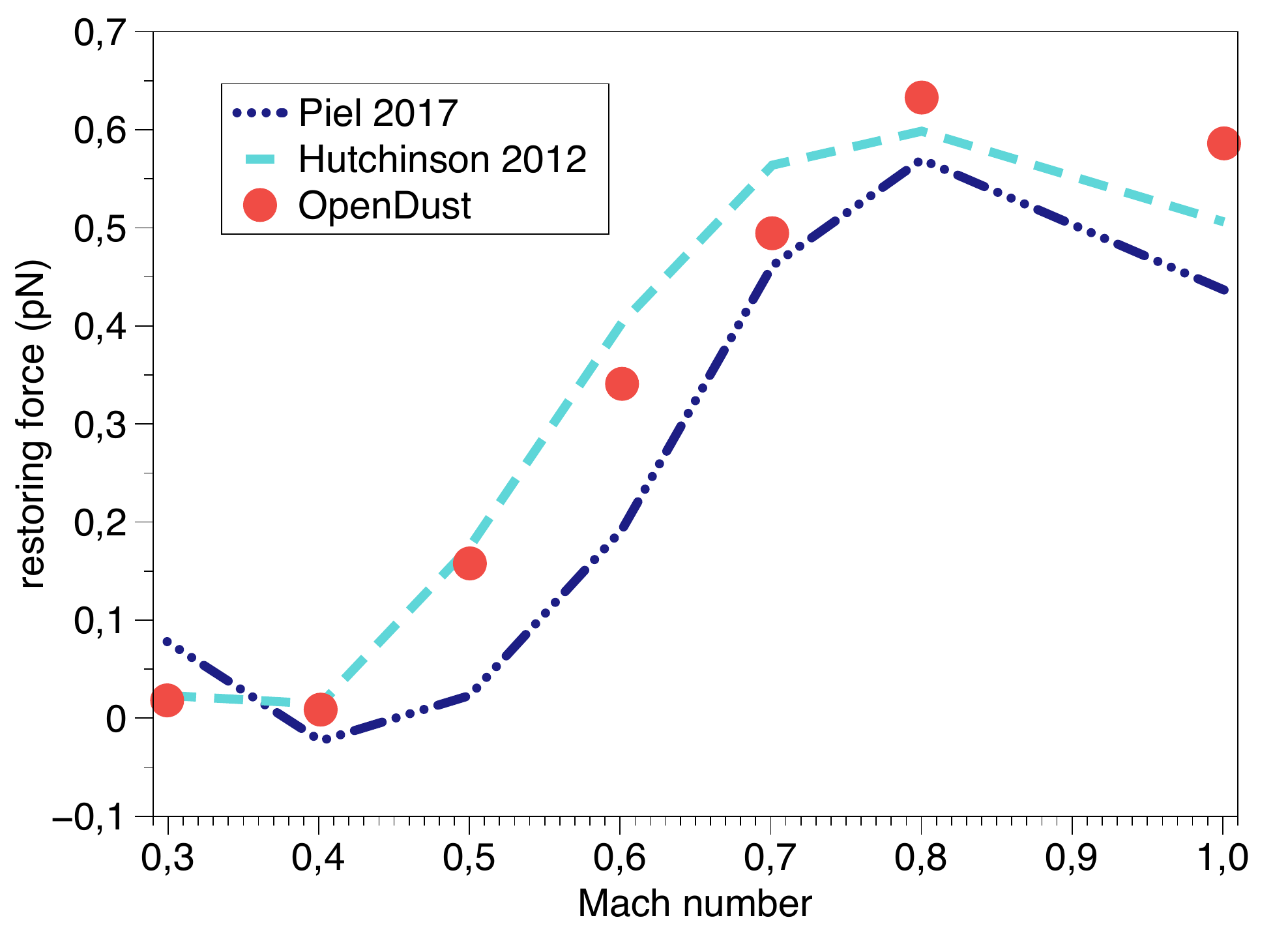}% 
    %\hspace{0.01 cm}
\caption{\label{fig:intergrainForces} Restoring horizontal force, acting on the lower dust particle. OpenDust results are depicted with red points, the light blue dashed line and the dark blue dashed-dotted line show results from
references \cite{hutchinson2012intergrain} and \cite{piel2017molecular} correspondingly.}
\end{figure}

The dependence of the restoring horizontal force, acting on the lower dust particle, on the ion flow Mach number is presented in the Figure \ref{fig:intergrainForces}. The OpenDust results are compared against previous calculations  \cite{piel2017molecular, hutchinson2012intergrain}. In the absence of a plasma flow, two similarly charged macroparticles would repel each other in the horizontal direction. However, the presence of a plasma flow leads to the fact that the lower particle is effectively attracted to the upper one, while the upper microparticle is repelled from the lower one. This effect is explained by focusing of ions behind the upper dust particle in the direction of the plasma flow. There is a good agreement between OpenDust and reference results.

\section{Performance}

In this section, we present OpenDust performance study. As a benchmark for OpenDust we use the first test-case from the Section 4 varying number of ions, number of integration time steps and size of the simulation domain. The bottleneck in OpenDust is the ion dynamic simulation, namely the inter-ion force calculation. Ion dynamics is performed in the \texttt{simulate} method of an \texttt{OpenDust} class object. For the performance testing, work time of this method is measured. The simulation time in OpenDust depends mostly on the number of used ions and integration time steps. Simulation time, obviously, is proportional to the number of integration time steps. This benchmark answers the question how the computational time changes with the variation of number of ions.   

In the first example from the Section 4, the cylindrical simulation domain of $6 r_{De}$ height and $3 r_{De}$ radius is used. Screened Coulomb interactions between almost all ions in such simulation domain can not be considered as negligible small. Thus, it is necessary to calculate pair forces between all ions every integration time step. The calculation algorithm for this calculation scales as $O(N^2)$, where $N$ is the number of ions or more accurately ion clouds. 

\begin{figure}
      \centering    \includegraphics[width=0.8\textwidth]{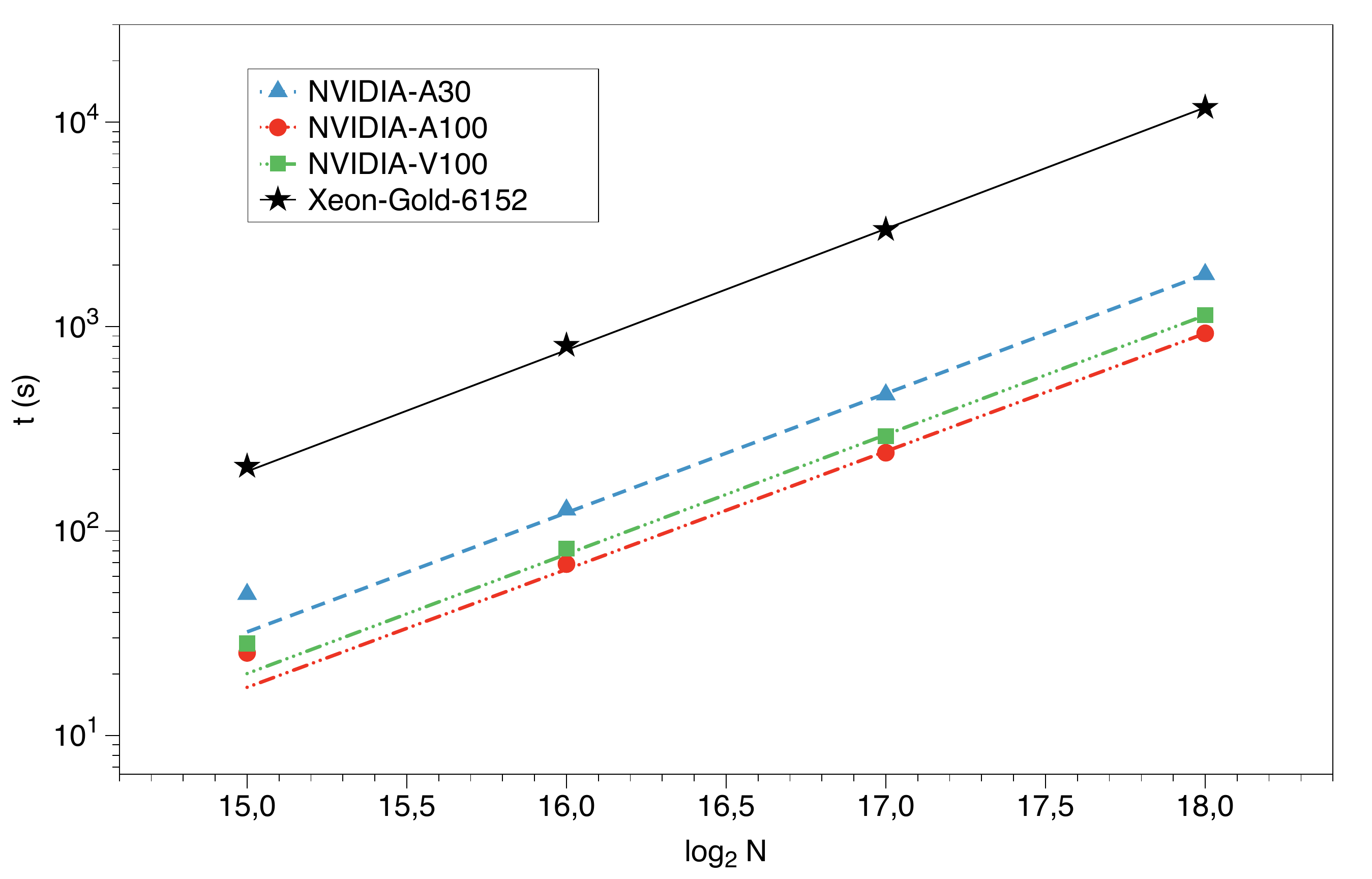}% 
    %\hspace{0.01 cm}
\caption{\label{fig:Performance} Mean execution time of the \texttt{simulate} routine for different number of ions. Measured time is depicted with symbols: NVIDIA A30 (blue triangles), NVIDIA Tesla V100 (green squares), NVIDIA Tesla A100 (red circles), CPU (black stars). Power function approximations of time values are presented with lines: NVIDIA A30 (blue dashed line), NVIDIA Tesla V100 (green dashed-triple-dotted line), NVIDIA Tesla A100 (red dashed-double-dotted line), CPU (black solind line). Obtained exponents for all cases are close to two. }
\end{figure}

As mentioned above, GPU-optimized OpenMM library \cite{eastman2017openmm} is used for numerical integration of the superions' equations of motion in OpenDust. Therefore, OpenDust inherits efficiency of OpenMM and can use GPUs for complex plasma modeling. The OpenDust version 1.0.0 uses Cuda core of OpenMM library and can be launched only on NVIDIA GPUs. In order to study OpenDust performance, we launch simulations corresponding to the first example from the Section 4 on different GPU cards: NVIDIA A30, NVIDIA Tesla V100 and NVIDIA Tesla A100. In addition, the simulation is launched using CPU implementation of OpenMM in order to compare CPU and GPU performance. For CPU calculations twenty two cores of  Intel Xeon Gold 6152 are used. Work time of \texttt{simulate} routine is measured for different number of ions 3000 integration time steps. In the Figure \ref{fig:Performance}, the dependence of the simulation time from number of ions $N$ is presented. Measured points are fitted with a power function $t(N) = \alpha N^\beta$, where $\alpha$ and $\beta$ are the fitting parameters. As expected obtained values for $\beta$ parameter are close to two, which corresponds to pair force calculation algorithm scaling $O(N^2)$.  CPU simulation time exceeds GPU simulation times approximately by one order of magnitude. NVIDIA Tesla A100 gives approximately two times better performance than NVIDIA A30 and $20\%$ better perfomance than NVIDIA Tesla V100. 

\begin{figure}
    \centering
    \includegraphics[width=0.8\textwidth]{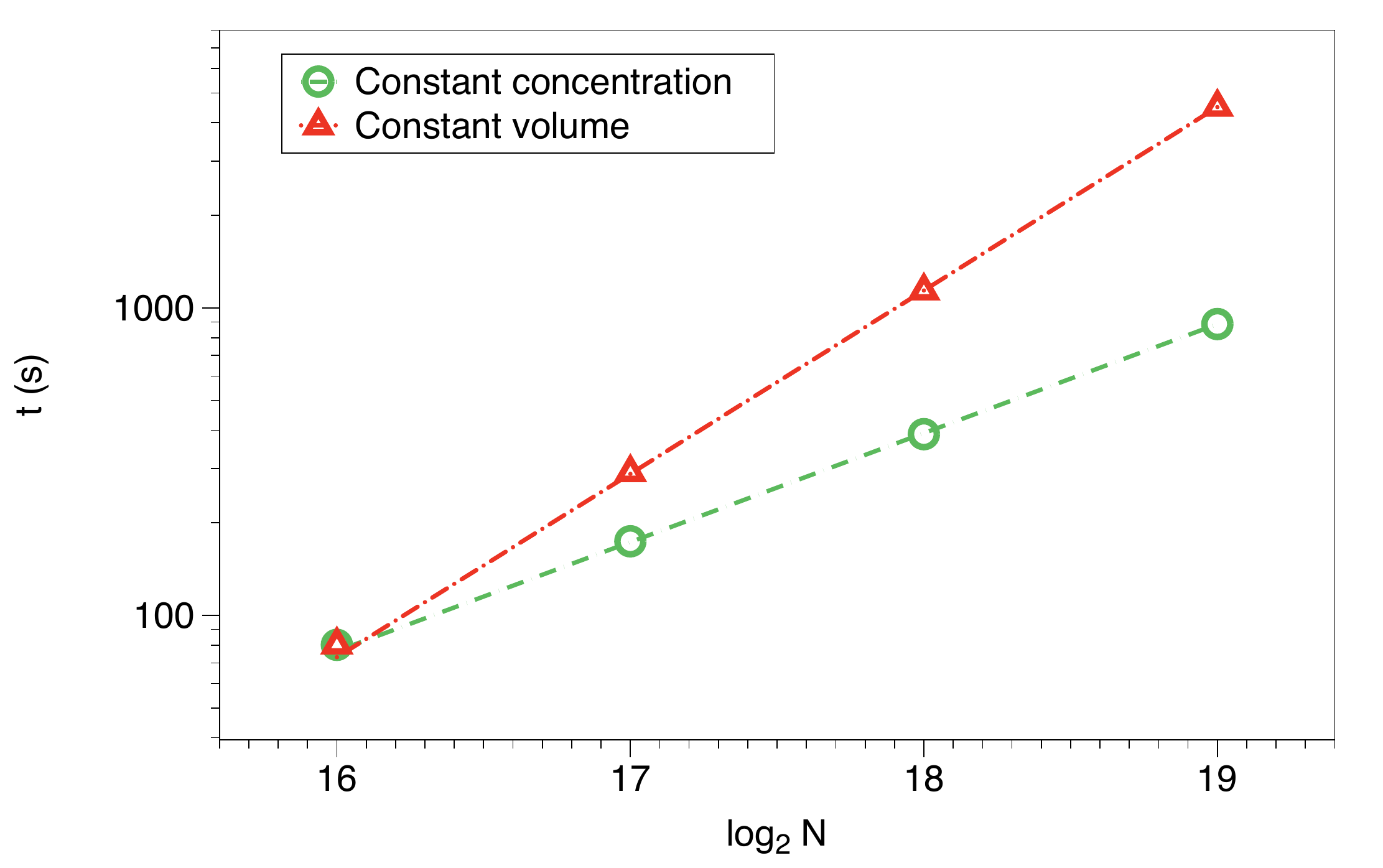}% 
    %\hspace{0.01 cm}
\caption{\label{fig:CutOff} dependence of simulation time on number of ions. Symbols denote measured times for constant volume simulation series (red triangle) and constant concentration series (green circles). Approximation of measured time points for constant volume and constant concentration series are depicted with red dashed-dotted line and green dashed line correspondingly.}
\end{figure}

When the subject of interest is a solitary dust particle, relatively small computational domain described above can be used in simulation. For simulation of a system with a greater number of dust particles the bigger computational domain may be necessary. The bigger domain means the rise of the number of ions in a simulation, which leads to a quadratic growth of the amount of interactions. However, when a computational domain gets bigger, larger number of ion-ion interactions, can be considered as negligible. In order to improve performance scaling for larger systems OpenDust uses “switching” function that is zero beyond a fixed cutoff distance. This allows to reduce scaling factor of algorithm complexity. We launch two series of plasma flow simulation around a solitary dust particle. First series is totally corresponds to the case used for previously discussed performance test. The size of the computational domain is constant and the number of ions is under variation. In the second series, the height of the cylindrical computational domain is changed with the number of ions to maintain constant concentration of ions. For calculations NVIDIA Tesla V100 is used. Measured time of each calculation is presented in the Figure \ref{fig:CutOff}. In the logarithmic scale, time points of two series lie on straight lines. The straight line for the first series corresponds to the quadratic dependence and the points of the second series are approximated with the following power function:
\begin{equation}
t(N) = 0.00016 N^{1.177},
\label{eq:cutOff}
\end{equation}
where N is the number of ions. Thus, using cutoff distance helps to reduce simulation complexity scaling on large systems almost up to linear dependence from number of ions.

\begin{figure}
  \begin{subfigure}{0.5\textwidth}
    \includegraphics[width=\linewidth]{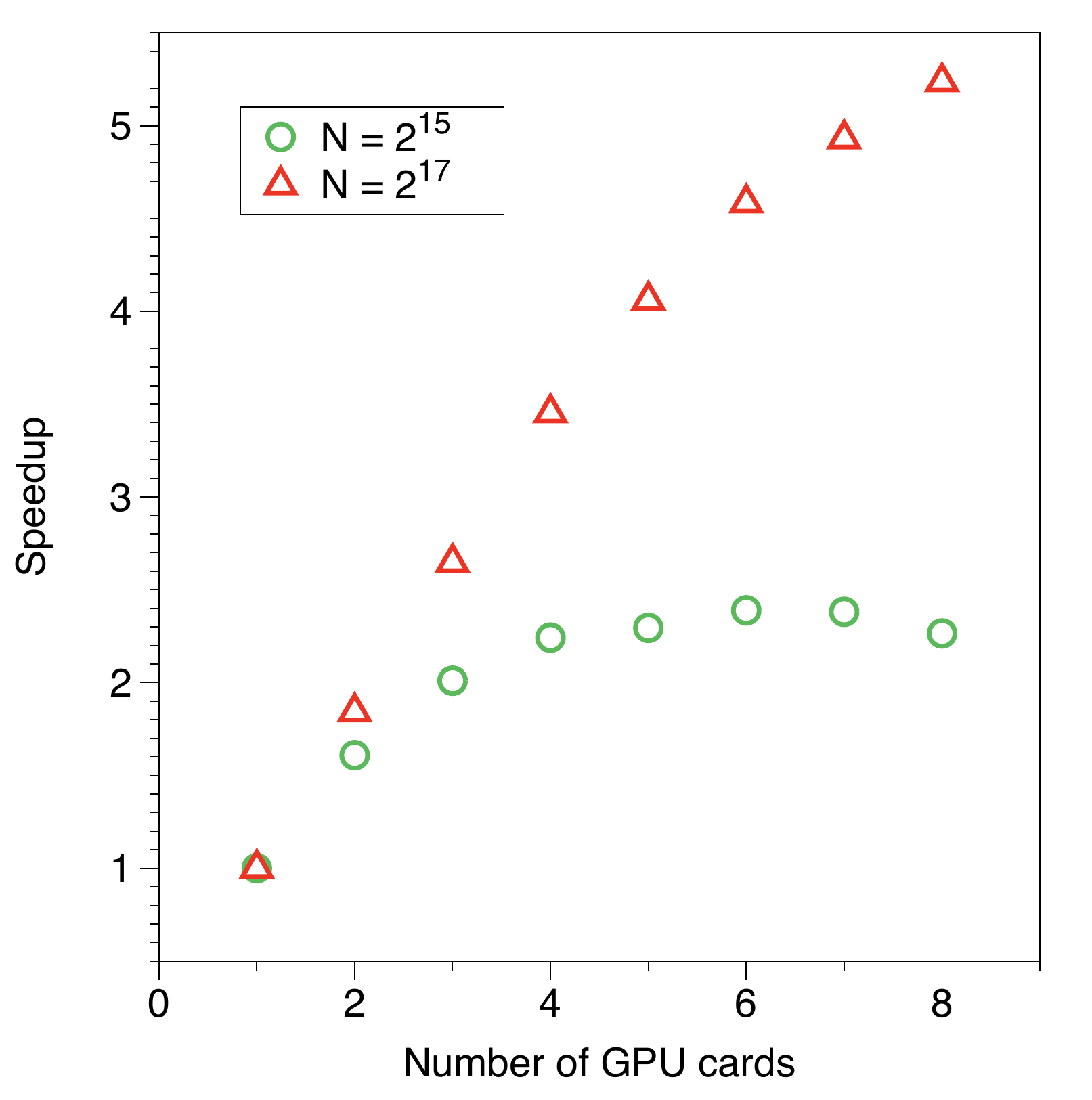}
    \caption{\label{fig:Speedup} Speedup}
  \end{subfigure}%
  \hspace*{\fill} 
  \begin{subfigure}{0.5\textwidth}
    \includegraphics[width=\linewidth]{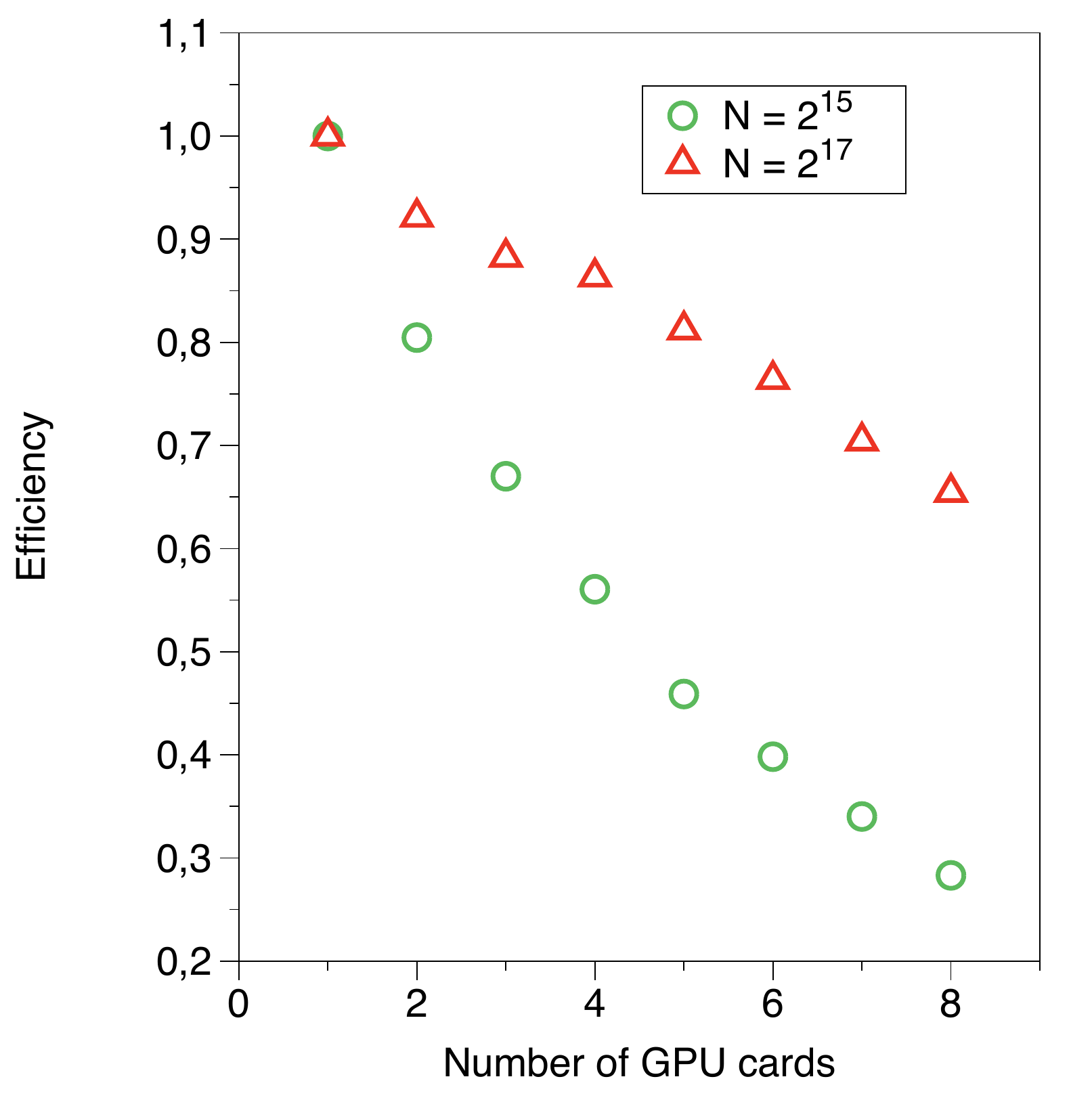}
    \caption{\label{fig:Efficiency} Efficiency}
  \end{subfigure}%
  \hspace*{\fill}
\caption{\label{fig:MultipleGPUs} The dependence of speedup and efficiency on the number of used GPU cards. Red triangles and green circles denote simulations with $2^{17}$ and $2^{15}$ ions correspondingly.}
\end{figure}

OpenDust allows launch simulation on multiple GPUs. Number of GPUs required for simulation can be set via keyword parameter \texttt{deviceIndex} of the \texttt{simulate()} routine. \texttt{deviceIndex} is a string with an enumeration of GPU devices indexes. For example, using of eight GPUs can be set with \texttt{deviceIndex = "0,1,2,3,4,5,6,7"}. In order to show performance growth with the number of using GPUs we measure simulation time for different number of GPUs and number of ions. In the Figure \ref{fig:MultipleGPUs}, speedup and efficiency of simulations with different number of GPUs are presented. For calculation NVIDIA Tesla A100 is used. Speedup and efficiency are measured for two numbers of ions $2^{15}$ and $2^{17}$. Speedup for $N = 2^{15}$ saturates at six GPUs and shows best speedup of value $~2.5$. Speedup for $N = 2^{17}$ does not reach saturation value even for eight GPUs. The reached value of speedup for  $N = 2^{17}$ is 5.5. Dependences in the Figure \ref{fig:Efficiency} shows that using of multiple GPUs for $N = 2^{17}$ is more efficient than for $N = 2^{15}$.

\section{Conclusions and outlook}

We have developed the first open-source GPU-based code for complex plasma modelling. OpenDust is provided as a Python library with an user-friendly interface and allows to calculate time-dependent forces, acting on dust particles, and dust charges in a plasma flow. The simulation launching process is divided into five logical steps: defining plasma, simulation and output parameters, defining dust particles and creating \texttt{OpenDust} class object. Developed interface allows users to set up simulation with a few lines of code. OpenDust is verified against previous results of complex plasma simulation and has shown good agreement with them. OpenDust makes it possible to self-consistently simulate plasma flow around dust particles just in seconds outperforming previous approaches of complex plasma simulation. In comparison, previously developed CPU-based code Coptic spends about an hour for such simulations \cite{hutchinson2011nonlinear}. OpenDust can effectively use multiple GPU cards which further speeds up the calculation. Moreover, OpenDust simulation complexity scales almost linearly with a growth of one dimension of computational domain and make it possible to simulate large dust particle systems which was unavailable before. OpenDust was developed to address the needs of computational theoretical and experimental complex plasma physicists. Furthermore, OpenDusts’ documentation provides a growing list of examples for common complex plasma physics problems; from a solitary dust particle to a cluster of interacting dust particles in a plasma flow.  

\section{Declaration of competing interest}
The authors declare that they have no known competing financial interests or personal relationships that could have appeared to influence the work reported in this paper.

\section{Acknowledgments}

This work was supported in part by the Ministry of Science and Higher Education of the Russian Federation (State Assignment No. 075-01056-22-00), in part within the framework of the HSE University Basic Research Program. This research was supported in part through computational resources of HPC facilities at HSE University and at JIHT RAS.

\newpage

\appendix
\section{}
Ions and dust particles in the computational domain are subject to a confinement force from the assumed infinite homogeneous distribution of ions outside the simulation region. The electric field from these ions is determined by first numerically calculating the shielded Coulomb potential of a homogeneously distributed ions in the cylindrical simulation domain. This potential is then subtracted from a constant uniform background potential, yielding the potential in a cylindrical cavity inside the homogeneous shielded Coulomb material \cite{matthews2020dust}. That potential is calculated at the beginning of the simulation on a sufficiently fine 2D grid 
\begin{equation}
\Phi_{ext}^{kp} = \Phi_0 - \frac{q_i}{4\pi\varepsilon_0}\sum_{l}\frac{1}{|\boldsymbol{r_i^l}-\boldsymbol{r_{grid}^{kp}}|}\exp{\left(-\frac{|\boldsymbol{r_i^l}-\boldsymbol{r_{grid}^{kp}}|}{r_{D_e}}\right)},
\label{eq:extPotential}
\end{equation}
\begin{equation}
\boldsymbol{r_{grid}^{kp}} = \left(\frac{k}{N_{grid}^x}R; 0; \frac{p}{ N_{grid}^p}\frac{H}{2}\right), 
\label{eq:grid}
\end{equation}
where $\Phi_0$ is the constant uniform background potential, $q_i$ is the charge of ions, $\boldsymbol{r_i^l}$ is the radius vector of the ion, $\boldsymbol{r_{grid}^{kp}}$ is the radius vector of the grid node, $k$ is the grid node index in the abscissa axis direction, $p$ is the grid node index in the applicate axis direction, $N_{grid}^x$ is the number of grid nodes in the abscissa axis direction, $N_{grid}^z$ is the number of grid nodes in the applicate axis direction.  
$\Phi_{ext}$ is then fitted with two dimensional eight-degree polynomial
\begin{equation}
\begin{aligned}
P(r,z) = \alpha_0 + \alpha_1 r^2 + \alpha_2 z^2 + \\ + \alpha_3 r^2 z^2 + \alpha_4 r^4+ \alpha_5 z^4 + \\ + \alpha_6 r^4 z^2 + \alpha_7 r^2 z^4 + \alpha_8 r^6+ \alpha_9 z^6 + \\ + \alpha_{10} r^4 z^4 + \alpha_{11} r^6 z^2 + \alpha_{12} r^2 z^6 + \alpha_{13} r^8 + \alpha_{14} z^8,
\label{eq:Polinomial}
\end{aligned}
\end{equation}
where $\alpha_0, ..., \alpha_{14}$ are the fitting parameters. 
The confining electric field is analytically calculated as a negative gradient of the fitted potential
\begin{equation}
\boldsymbol{E_{ext}(x,y,z)} = \left(-\frac{\partial P}{\partial r}\frac{x}{\sqrt{x^2+y^2}}; -\frac{\partial P}{\partial r}\frac{y}{\sqrt{x^2+y^2}}; -\frac{\partial P}{\partial z}\right). 
\label{eq:extField}
\end{equation}

\section{}
Assume the ion flow through the cylindrical computational domain is stationary and homogeneous with ion velocity distribution function
\begin{equation}
f(v_x,v_y,v_z) = \varphi_M(v_x)\varphi_M(v_y)\varphi(v_z),
\label{eq:distribution}
\end{equation}
where $v_x, v_y, v_z$ are the velocities in Cartesian coordinates, $\varphi_M(v_x)$ is the Maxwell distribution for $v_x$, $\varphi_M(v_y)$ is the Maxwell distribution for $v_y$ and $\varphi(v_z)$ is the distribution for $v_z$. The integral flow of ions, entering computational domain, through cylinder's boundary then can be written in the following form:
\begin{equation}
J_{bottom} = \pi R^2 \int_{0}^{\infty} v_z \varphi(v_z)d v_z,
\label{eq:integralFlowBottom}
\end{equation}
\begin{equation}
J_{top} = -\pi R^2 \int_{-\infty}^{0} v_z \varphi(v_z) d v_z,
\label{eq:integralFlowTop}
\end{equation}
\begin{equation}
J_{side} = 2\pi R L \int_{0}^{\infty} v_r \varphi_M(v_r) d v_r,
\label{eq:integralFlowSide}
\end{equation}
where $J_{bottom}, J_{top}, J_{side}$ are the integral flows of ions, entering the computational domain trough the bottom, top and side of the cylinder correspondingly. 

In the OpenDust boundary conditions algorithm, any ion, that leaves the computational domain or is absorbed by dust particles, is replaced by a newly injected ion at the bottom, top or side boundary of the cylinder with the following probabilities:
\begin{equation}
p_{bottom} = \frac{J_{bottom}}{J_{bottom}+J_{top}+J_{side}},
\label{eq:probabilityBottom}
\end{equation}
\begin{equation}
p_{top} = \frac{J_{top}}{J_{bottom}+J_{top}+J_{side}},
\label{eq:probabilityTop}
\end{equation}
\begin{equation}
p_{side} = \frac{J_{side}}{J_{bottom}+J_{top}+J_{side}}.
\label{eq:probabilitySide}
\end{equation}
If an ion is newly injected on the top or bottom boundary, then $v_x$ and $v_y$ are chosen from Maxwell distribution and $v_z$ is chosen from the cumulative distribution for the bottom,
\begin{equation}
G_{bottom} = \int_{0}^{v_z} v_z^{'} \varphi(v_z^{'})d v_z^{'},
\label{eq:cummulativeDistributionBottom}
\end{equation}
and for the top,
\begin{equation}
G_{top} = -\int_{v_z}^{0} v_z^{'} \varphi(v_z^{'}) d v_z^{'},
\label{eq:cummulativeDistributionTop}
\end{equation}
boundaries correspondingly. If an ion is newly injected on the side boundary, then $v_z$ is generated according to the $\varphi(v_z)$ distribution and $v_x$ and $v_y$ are chosen using the following formula:
\begin{equation}
v_x = -\frac{x}{\sqrt{x^2+y^2}}v_r,
\label{eq:vx}
\end{equation}
\begin{equation}
v_y = -\frac{y}{\sqrt{x^2+y^2}}v_r,
\label{eq:vy}
\end{equation}
where $v_r$ is the radial velocity randomly chosen according to the following cumulative distribution function:
\begin{equation}
G_{side} = \int_{0}^{v_r} v_r^{'} \varphi_M(v_r^{'}) d v_r^{'}.
\label{eq:cummulativeDistributionSide}
\end{equation}



%% References


\bibliographystyle{elsarticle-num}
\bibliography{library.bib}






% \end{thebibliography}

\end{document}